\documentclass{aastex631}
\usepackage{amsmath}
\usepackage{graphicx} 

\begin{document}

\title{Machine Learning-Driven Analysis of kSZ Maps to Predict CMB Optical Depth $\tau$}

\correspondingauthor{Farshid Farhadi Khouzani}
\email{farshid.farhadikhouzani@unlv.edu}

\author[0009-0003-8283-8391]{Farshid Farhadi Khouzani}
\author[0000-0002-6123-4383]{Abinash Kumar Shaw}
\affiliation{Department of Computer Science, University of Nevada, Las Vegas, NV 89154, USA}

%\collaboration{20}{(PASP Editor-in-Chief)}

\author[0000-0002-4693-0102]{Paul La Plante}
\affiliation{Department of Computer Science, University of Nevada, Las Vegas, NV 89154, USA}
\affiliation{Nevada Center for Astrophysics, University of Nevada, Las Vegas, NV 89154, USA}
\author[0009-0006-2124-1032]{Bryar Mustafa Shareef}
\author{Laxmi Gewali}
\affiliation{Department of Computer Science, University of Nevada, Las Vegas, NV 89154, USA}



\begin{abstract}

Upcoming measurements of the kinetic Sunyaev-Zel'dovich (kSZ) effect, which results from Cosmic Microwave Background (CMB) photons scattering off moving electrons, offer a powerful probe of the Epoch of Reionization (EoR). The kSZ signal contains key information about the timing, duration, and spatial structure of the EoR. A precise measurement of the CMB optical depth $\tau$, a key parameter that characterizes the universe's integrated electron density, would significantly constrain models of early structure formation. However, the weak kSZ signal is difficult to extract from CMB observations due to significant contamination from astrophysical foregrounds. We present a machine learning approach to extract $\tau$ from simulated kSZ maps. We train advanced machine learning models, including swin transformers, on high-resolution seminumeric simulations of the kSZ signal. To robustly quantify prediction uncertainties of $\tau$, we employ the Laplace Approximation (LA). This approach provides an efficient and principled Gaussian approximation to the posterior distribution over the model's weights, allowing for reliable error estimation. We investigate and compare two distinct application modes: a post-hoc LA applied to a pre-trained model, and an online LA where model weights and hyperparameters are optimized jointly by maximizing the marginal likelihood. This proof-of-concept approach provides a framework for constraining $\tau$ and its associated uncertainty, which could be adapted for analysis of upcoming CMB surveys like the Simons Observatory and CMB-S4.

\end{abstract}



\section{Introduction} \label{sec:intro}

One of the most significant transformations in cosmic history is the Epoch of Reionization (EoR), the period when the first generation of stars and galaxies formed, emitting radiation that ionized the vast reservoirs of neutral hydrogen in the intergalactic medium (IGM). This event marked the end of the cosmic ``dark ages'' and fundamentally reshaped the universe into the ionized state we observe today. A detailed understanding of the EoR is critical for constraining models of early galaxy formation, the properties of the first luminous sources, and the evolution of the IGM \citep{2001PhR...349..125B,2006PhR...433..181F,2013fgu..book.....L}.

%Observational efforts to probe this era primarily focus on two key signatures: the redshifted 21 cm line from neutral hydrogen and secondary anisotropies in the Cosmic Microwave Background (CMB). While upcoming radio interferometers like HERA and the SKA aim to map the 21 cm signal, this approach faces the immense challenge of disentangling the faint cosmological signal from astrophysical foregrounds that are orders of magnitude brighter.

One observational probe of the EoR is the kinetic Sunyaev-Zel'dovich (kSZ) effect \citep{1972CoASP...4..173S}. The kSZ signal is a secondary anisotropy of the cosmic microwave background (CMB) created when CMB photons inverse-Compton scatter off free electrons in ionized regions (or ``bubbles'') that have a peculiar velocity relative to the CMB rest frame. The resulting temperature fluctuations directly trace the distribution and motion of ionized gas, providing a unique window into the morphology and dynamics of reionization. Upcoming and ongoing CMB experiments, such as the Simons Observatory \citep{2019JCAP...02..056A}, are designed to make high-fidelity measurements of this signal. %Because the kSZ effect is observed at microwave frequencies and has different foreground contaminants (e.g., the thermal SZ effect and cosmic infrared background) than the 21 cm signal, it offers a powerful, independent pathway to studying the EoR with largely uncorrelated systematic errors.

Using measurements of the EoR, it is possible to infer key properties of this epoch, such as its timing and duration. These properties are indirectly encoded in the CMB optical depth to reionization $\tau$. This parameter, which measures the integrated column density of free electrons along the line of sight, is one of the fundamental parameters of the standard $\Lambda$CDM cosmological model. However, current constraints on $\tau$ from CMB experiments like \textit{Planck} still carry significant uncertainty ($\tau=0.054 \pm 0.007$, \citealt{2020A&A...641A...6P}), which in turn limits the precision of other key parameters like $\Omega_m$ and $\sigma_8$. A more precise and robust measurement of $\tau$ would therefore have a profound impact on cosmology, most directly in the determination of the amplitude of the primordial power spectrum $A_S$ \citep{2016PhRvD..93d3013L}.

Extracting the faint, non-Gaussian kSZ signal from CMB maps is a significant data analysis challenge. Measurements from the South Pole Telescope show a $\sim$3$\sigma$ detection of the reionization-era kSZ signal \citep{2021ApJ...908..199R}, which places some constraints on the ionization history of the Universe. Traditional methods often rely on statistical estimators like the power spectrum, which are insensitive to the rich morphological information contained in the maps. An alternative and promising approach is to apply supervised machine learning techniques directly to images of the CMB. One such study has shown the potential of machine learning for analyzing CMB data, for example, by using neural networks for full-sky foreground cleaning \citep{2020ApJ...903..104P}. Other studies have demonstrated the power of convolutional neural networks (CNNs) being used to successfully extract $\tau$ from simulated 21 cm maps \citep{2021PASP..133d4001B,2022PASP..134d4001Z}. The work presented here goes beyond this previous work by using vision transformers, an attention-based technique for performing image-based tasks in computer vision like classification, regression, or semantic segmentation. This study also includes a method for estimating the full posterior distribution of a regression parameter instead of a single point-estimate, which is the typical output of conventional neural networks.

In this paper, we build upon and advance this machine learning paradigm. We apply a shifted window (swin) transformer \citep{2021arXiv210314030L} to simulated kSZ maps to perform a regression for the value of $\tau$. A key innovation of our work is the method used for uncertainty quantification. To move beyond a single point estimate and generate a well-calibrated error bar, we employ the Laplace Approximation (LA, \citealt{2021arXiv210614806D}). We explore two distinct schemes for its application: a post-hoc approach, where the LA is applied to a fully trained network, and an online approach, where the model and hyperparameters are trained jointly by optimizing the LA to the marginal likelihood. This comparison allows us to assess the trade-offs between leveraging a pre-trained, high-performing model versus a fully Bayesian optimization scheme. Both approaches provide a principled and computationally inexpensive framework for estimating the uncertainty in our predictions, marking a significant step forward in applying machine learning to cosmological inference.

This paper is organized as follows: In Section 2, we describe the semi-numeric simulations used to generate our kSZ maps and the corresponding ground-truth $\tau$ values. In Section 3, we detail our machine learning architectures, training methodology, and the implementation of our Bayesian layers. In Section 4, we present the results of our regression and the performance of our uncertainty quantification. In Section 5, we discuss the interpretation of our results and compare our forecasted constraints with those from other methods. We also discuss potential extensions of our work in future studies.

\section{Reionization Simulations} \label{sec:data}

In this section, we describe the simulation methods used to model the EoR and the kSZ field. Given the angular resolution of current and upcoming CMB observations, we opt to simulate relatively large volumes with moderate resolution, as this helps ensure a statistically significant measurement for the scales of interest. We begin by describing the seminumeric simulation methods used in our study, as well as a description of the generation of the kSZ field.

\subsection{EoR Modeling}
Accurate numeric simulations of the EoR are an extraordinarily difficult computational task. Although it is possible to run fully coupled simulations that tracks to co-evolution of dark matter, baryons, and photons simultaneously, even modern supercomputers do not allow for simulating volumes much larger than $\sim 100$ $h^{-1}$Mpc. Given that we are interested in simulating the kSZ signal as a way of inferring the value of $\tau$, we require resolving sufficiently large volumes ($\gtrsim 1$ $h^{-1}$Gpc) that cosmic variance does not affect the result. Additionally, the angular scales probed by current- and next-generation CMB telescopes correspond to roughly $1000 \lesssim \ell \lesssim 6000$, which means that small-scale features of high-resolution simulations would be lost observationally. Thus, we opt to use a seminumeric scheme for simulating reionization, where we are able to capture the large-scale features of the EoR reasonably well while still remaining computationally efficient enough to run multiple realizations of large volumes. This latter requirement is important for generating a sufficiently large training set for machine learning applications, including having different ionization histories and, by extension, different $\tau$ values.

We make use of the \texttt{zreion} seminumeric code \citep{2013ApJ...776...81B}, which has been used in previous studies of the kSZ signal from the EoR \citep{2013ApJ...776...83B,2020ApJ...899...40L,2022ApJ...928..162L,2025ApJ...991..195Z}, as well as machine learning-based parameter inference studies \citep{2019ApJ...880..110L,2021PASP..133d4001B,2022PASP..134d4001Z}. The central \textit{Ansatz} of \texttt{zreion} is that the dark matter field $\delta_m(\boldsymbol{r})$ and the redshift at which a particular portion of the IGM is reionized $z_\mathrm{re}(\boldsymbol{r})$ are correlated on large scales. This result is generally true for inside-out reionization scenarios, where the densest regions of the Universe ionize first. We begin by defining the matter overdensity field $\delta_m(\boldsymbol{r})$ as:
\begin{equation}
\delta_m(\boldsymbol{r}) \equiv \frac{\rho_m(\boldsymbol{r}) - \bar{\rho}_m}{\bar{\rho}_m},
\label{eqn:delta_m}
\end{equation}
where $\bar{\rho}_m$ is the mean matter density. We define a corresponding ``overdensity'' field for the redshift of reionization $\delta_z(\boldsymbol{r})$, such that:
\begin{equation}
\delta_z(\boldsymbol{r}) \equiv \frac{\left[z_\mathrm{re}(\boldsymbol{r}) + 1\right] - \left[\bar{z} +1\right]}{\bar{z} + 1},
\end{equation}
where $\bar{z}$ is the average redshift of the field $z_\mathrm{re}(\boldsymbol{r})$. The \texttt{zreion} method posits that the relationship between $\delta_m(\boldsymbol{r})$ and $\delta_z(\boldsymbol{r})$ can be expressed as a scale-dependent bias factor in Fourier space $b_{zm}(k)$. Specifically:
\begin{equation}
b_{zm}^2(k) \equiv \frac{\left\langle\delta^*_z \delta_z\right\rangle_k}{\left\langle\delta^*_m\delta_m\right\rangle_k} = \frac{P_{zz}(k)}{P_{mm}(k)},
\end{equation}
where $P_{xx}(k)$ is the three-dimensional spherically-averaged auto-power spectrum of the field $\delta_x$. This bias factor $b_{zm}$ is parametrized by three parameters: $b_0$, $\alpha$, and $k_0$. The bias can be expressed in Fourier space as a function of spherical wavenumber $k$:
\begin{equation}
b_{zm}(k) = \frac{b_0}{\left(1 + \frac{k}{k_0}\right)^\alpha}.
\end{equation}
We fix the value of $b_0 = 1 / \delta_c = 0.593$, where $\delta_c$ is the critical overdensity in spherical collapse halo models. Thus, the model only depends on the set of parameters $\{ \bar{z}, \alpha, k_0 \}$. Given a dark matter density field and a choice of these parameters, it is possible to calculate the redshift of reionization field $z_\mathrm{re}(\boldsymbol{r})$. By extension, the ionization state of a given voxel of the IGM can be computed at a specific redshift $z_0$: if $z_\mathrm{re}(\boldsymbol{r}) \geq z_0$, the voxel is assumed to be totally ionized, and totally neutral if $z_\mathrm{re}(\boldsymbol{r}) < z_0$.

\subsection{kSZ Map Generation}
\label{sec:ksz_maps}

To generate the kSZ maps used for this study, we run a series of dark-matter-only simulations that contain 1024$^3$ particles in a cubic comoving volume with a length of $L = 2\; h^{-1}\mathrm{Gpc}$ on a side, which corresponds roughly to an angular extent of $\theta \approx 20^\circ$ at $z = 6$. We apply \texttt{zreion} to the resulting volume to calculate the redshift of reionization field $z_\mathrm{re}(\boldsymbol{r})$, which allows for calculating the free electron number density $n_e$ for a given point in the volume at any specific redshift $z_0$. The kSZ effect comes from CMB photons inverse Compton scattering off of free electrons in the IGM moving with a peculiar velocity relative to the observer $\boldsymbol{v}\cdot \hat{n}$, where $\hat{n}$ is the line-of-sight vector of the observer. The change in the observed temperature of the CMB is given by \citep{1972CoASP...4..173S}:
\begin{equation}
\frac{\Delta T(\hat{n})}{T_\mathrm{CMB}} = -\int\mathrm{d}\chi\, g(\chi)\, \boldsymbol{q} \cdot \hat{n},
\label{eqn:ksz}
\end{equation}
where $\chi$ is the comoving distance along the line-of-sight and $\boldsymbol{q}$ is the local electron momentum field: $\boldsymbol{q} = \boldsymbol{v}(1 + \delta_m)(1 + \delta_x)/c$. In this expression, $\delta_x$ is the local ionization overdensity, defined analogously to the matter overdensity in Equation~(\ref{eqn:delta_m}). The quantity $g(\chi)$ is the visibility function, which quantifies the probability that a CMB photons scatters between $\chi$ and $\chi - \mathrm{d}\chi$ without subsequent scattering along the path to the observer. This can be written as \citep{2016ApJ...824..118A}:
\begin{equation}
g(\chi) = \frac{\partial \left[e^{-\tau(\chi)}\right]}{\partial \chi} = e^{-\tau(\chi)} \sigma_T n_{e,0} \left\langle x_i \right\rangle (1 + z)^2,
\label{eqn:visibility}
\end{equation}
where $\sigma_T$ is the cross-section for Thomson scattering, $\left\langle x_i \right\rangle$ is the globally averaged ionization fraction, and $n_{e,0} = [1 - (4 - N_\mathrm{He})Y/4]\Omega_b \rho_\mathrm{crit}/m_p$ is the mean electron number density at $z=0$. Here, $\tau(\chi)$ is the electron-scattering optical depth between the observer and the location represented by a comoving distance $\chi$. Although in principle $e^{-\tau(\chi)}$ must be computed along each line of sight for each comoving distance $\chi$, we instead opt to use the globally averaged value of $\left\langle x_i (z) \right\rangle$ to compute a value for $\tau(\chi)$. This simplification is justified because the probability of a scattering event is small, and the variation between sightlines is a relatively unimportant effect. To include the effect of helium \textbf{reionization}, we set $N_\mathrm{He} = 1$, which assumes that helium is singly ionized when hydrogen is ionized. Helium is widely thought to be doubly ionized at later redshifts after a significant increase in quasar activity \citep{2017ApJ...841...87L}. We use Equations~(\ref{eqn:ksz}) and (\ref{eqn:visibility}) to generate the kSZ map given a simulation volume.

We construct our kSZ maps in the following way. After using \texttt{zreion} as described above, we compute kSZ sightlines that trace plane-parallel lines through the volume using the flat-sky approximation. This grid of sightlines is generated for fixed angular coordinates $(\theta_x, \theta_y)$, and so an interpolation must be done from the fixed comoving coordinates of the simulation volume to the sightline coordinates. To compute the local dark matter density and velocity values, which are needed for the electron momentum at a particular location and redshift $\boldsymbol{q}(\boldsymbol{x}, z)$, we generate second-order Lagrangian perturbation theory (2LPT) snapshots at bracketing redshift values $z_i$ and $z_{i+1}$ such that $z_i \leq z < z_{i+1}$. We then linearly interpolate in scalefactor $a = 1/(1+z)$ to find the matter density and velocities at the desired redshift. The ionization state of the voxel is given by the redshift of reionization field $z_\mathrm{re}(\boldsymbol{r})$. We then sum up the contribution along each line of sight to obtain a two-dimensional kSZ map.

Note that these maps, by construction, only contain the contribution from the reionization-era, so-called ``patchy'' kSZ, and not the late-time homogeneous kSZ. In future work, we plan to account for observational effects and contamination from competing signals, like the late-time kSZ, though for the current work we use these ``pristine'' maps of the reionization-era kSZ as a proof-of-concept. Real CMB data would of course have contributions from other components not relevant to reionization, such as the late-time kSZ, the (lensed) primary CMB, the thermal Sunyaev-Zel'dovich (tSZ) effect, the cosmic infrared background (CIB), foregrounds, and instrumental effects from real instruments. In principle, one could apply a Wiener filter constructed for the purpose of isolating this patchy-kSZ signal from CMB data, as in \cite{2022ApJ...928..162L} or \cite{2025ApJ...991..195Z}. Such a filter could be applied in harmonic (Fourier) space to data which has been Fourier transformed, and then inverse Fourier transformed back to real (map) space. Although the kSZ maps would be visually quite different than the pristine maps, in principle the relevant information would be preserved. Any neural networks would have to be retrained using these Wiener filtered maps as input data, up to and possibly including hyperparameter tuning (discussed more below in Sec.~\ref{subsec:hyperparam}). We save such an investigation for future work.

\begin{figure}[t]
 \centering
 \includegraphics[width=0.7\columnwidth]{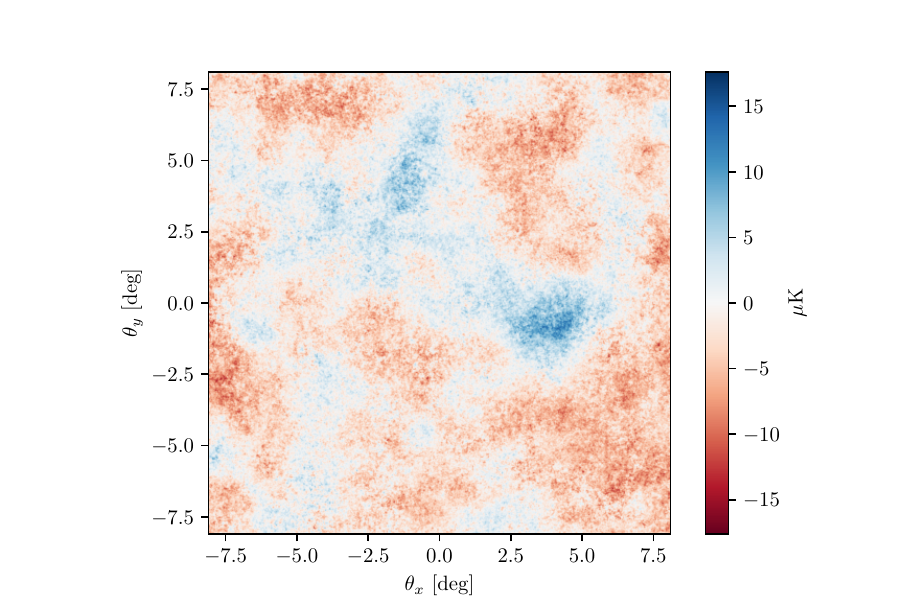}
 \caption{An example of a simulated kSZ map used in this work. The colors represent the temperature fluctuation in the CMB in units of microkelvin ($\mu$K). These fluctuations are caused by the scattering of CMB photons off of ionized bubbles moving with a peculiar velocity during the Epoch of Reionization. This map serves as a single input image for our machine learning models.}
 \label{fig:ksz_map}
\end{figure}

Following the procedure outlined above and using Equation~(\ref{eqn:ksz}), we generate the two-dimensional kSZ maps that serve as the input for our machine learning models. An example of one such simulated map, corresponding to a single reionization history, is shown in Figure \ref{fig:ksz_map}. The map displays the temperature fluctuations in microkelvin ($\mu$K) caused by CMB photons scattering off moving ionized bubbles. Hot spots (positive $\mu$K) correspond to regions where the ionized gas has a net peculiar velocity toward the observer, while cold spots (negative $\mu$K) correspond to gas moving away. The complex, non-Gaussian structure of these features contains the morphological information from which our network will learn to infer the integrated optical depth $\tau$. Each map generated for our training, validation, and test sets has a corresponding ground-truth $\tau$ value calculated from its unique reionization history.

Throughout our kSZ map creation process, we employ the flat-sky approximation to generate non-periodic ``squares'' of patches of the simulated CMB sky. In practice, real data could include a full-sky map with periodic boundaries as a result of the curvature of the sky. As discussed more below in Sec.~\ref{subsec:network}, the initial stage of our vision transformer uses a shifted window that naturally adapts to periodicity in input images. However, the technique assumes rectilinear pixels in the input image, which is not necessarily true for pixelization schemes of the sphere (e.g., HEALPix). As such, a reprojection to a regular grid would almost certainty be required to make use of the approach outlined below.

% Notes for Figure 1:
%  - use a different font so that it matches more closely to the text of the manuscript:
%     plt.rcParams.update({"text.usetex": True, "font.family": "serif"})
%  - change the x- and y-axis labels to be in angular units (degrees) rather than pixel values:
%    from astropy.cosmology import Planck18
%    from astropy import units
%    lbox = 2 * units.Gpc / Planck18.h
%    z = 6.0  # redshift 6
%    a = 1 / (1 + z)  # get scalefactor
%    r_phys = a * lbox
%    theta = float(r_phys / Planck18.angular_diameter_distance(z))
%    theta_deg = theta * 180 / np.pi
%    # then, in the call to imshow, you should set something like:
%    ax.imshow(..., extent=[-theta_deg / 2, theta_deg / 2, -theta_deg / 2, theta_deg / 2])
%    ax.set_xlabel(r"$\theta_x$ [deg]")
%    ax.set_ylabel(r"\theta_y$ [deg]")

\subsection{Dataset Creation}

In order to generate a dataset suitable for training and testing a machine learning network, we must have a sufficiently large number of realizations using different combinations of \texttt{zreion} parameters to span the plausible space of $\tau$ values allowed by current observations. To this end, we perform 1,000 simulations with different combinations of the \texttt{zreion} parameters $\{\bar{z}, \alpha, k_0\}$. Each of these simulations features a different set of initial conditions, though with cosmological parameters fixed to those of the most recent results of \textit{Planck} \citep{2020A&A...641A...6P}.

Given a realization of initial conditions plus a particular choice of \texttt{zreion} parameters, we generate the kSZ map using the methods outlined above in Sec.~\ref{sec:ksz_maps}. We record the corresponding value for $\tau$ by directly integrating the ionization history of the volume. Thus, we have 1,000 samples containing kSZ maps and matching values of $\tau$. In our approach outlined below, we use the kSZ maps as the input data for our machine learning network and the value of $\tau$ as the output. We enforced strict reproducibility across all experiments by fixing the random seed to 42 for weight initialization, data shuffling, and other stochastic processes. The 1,000 generated kSZ maps were partitioned deterministically into a training set ($70\%$, 700 maps), a validation set ($15\%$, 150 maps), and a held-out test set ($15\%$, 150 maps). This rigorous splitting strategy ensured that the test set served as an independent evaluation metric, having been excluded from all training and hyperparameter tuning phases.

\begin{figure}[ht!]
    \centering
    \includegraphics[width=0.7\linewidth]{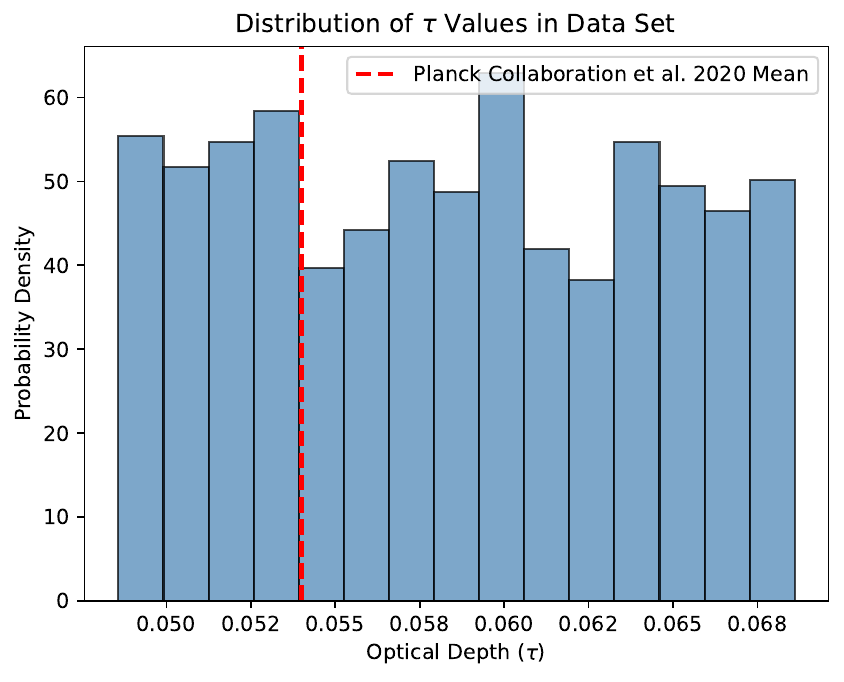}
    \caption{The normalized probability density of the CMB optical depth $\tau$ across the entire dataset of 1,000 simulations generated for this study. The distribution (blue bars) is centered near $\tau \approx 0.06$ and fully encompasses the central value of the \citealt{2020A&A...641A...6P} constraint ($\tau = 0.054$, red dashed line). This confirms that the full dataset spans a physically comprehensive parameter space prior to splitting into training, validation, and testing sets.}
    \label{fig:tau_dist}
\end{figure}

To ensure our dataset covers the physically viable parameter space, we visualize the distribution of the optical depth $\tau$ across our 1,000 simulations in Figure~\ref{fig:tau_dist}. The distribution spans the range $\tau \in [0.048, 0.069]$, peaking near $\tau \approx 0.06$. Crucially, this range encompasses the current best-fit value from \citealt{2020A&A...641A...6P} ($\tau = 0.054 \pm 0.007$, indicated by the red dashed line), as well as a range of values above and below this quantity. This confirms that our semi-numeric approach captures a diverse set of reionization histories consistent with standard cosmological models, allowing the network to learn from a representative sample of potential physical realities.

\section{Machine Learning Methods} \label{sec:model}

\begin{figure}[t]
  \centering  
  %\includesvg[width=0.8\textwidth]{nn}
  \includegraphics[width=\textwidth]{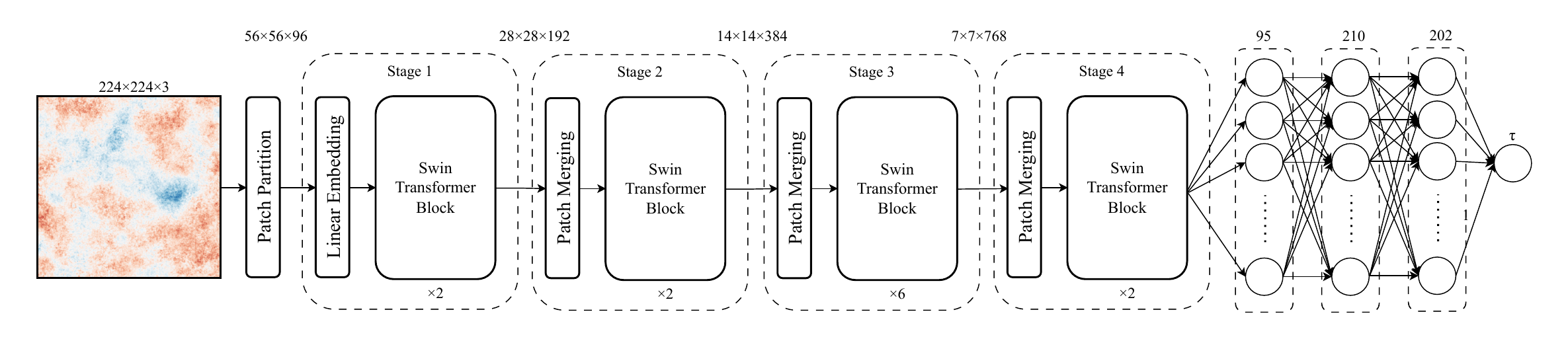}
  % \caption{A visualization of the hybrid Swin Transformer architecture used in our analysis. The large cubes represent the main stages of the Swin Transformer backbone, which processes the input image. The feature vector is then passed to a regression head composed of fully-connected dense layers (depicted as vertical bars). The Laplace Approximation is applied post-hoc to this trained network to infer a posterior distribution over the weights. The input images have dimensions of $224 \times 224 \times 3$, and the network's final output is a single value for the optical depth, $\tau$. The figure was generated using the NN-SVG tool.}
  \caption{The end-to-end architecture of the Swin Transformer model used in this work for regressing the optical depth, $\tau$. An input kSZ map is first preprocessed to a size of $224 \times 224$ pixels. The image is divided into non-overlapping $4 \times 4$ patches and linearly embedded into a 96-dimensional feature space. This is followed by a four-stage Swin Transformer backbone, where patch merging layers progressively downsample the spatial resolution (from $56 \times 56$ to $7 \times 7$) while increasing the feature dimension (from 96 to 768). The output feature vector is then passed to a Multi-Layer Perceptron (MLP) regression head. The dimensions of the three hidden layers (95, 210, and 202 neurons) were determined via the hyperparameter optimization described in Section \ref{subsec:hyperparam}, selected from a search space of [16, 256] neurons. The final output neuron produces the point estimate for $\tau$. For the post-hoc configuration, this entire trained model is passed to the Laplace Approximation library to compute a posterior distribution over the MLP head weights.}
  \label{fig:my_svg}
\end{figure}

In this section, we describe the machine learning framework used to infer the CMB optical depth, $\tau$, from simulated kSZ maps. Our approach is designed not only to achieve high predictive accuracy but also to provide robust, principled uncertainty estimates for our predictions. To accomplish this, we develop a hybrid architecture that leverages the feature extraction power of a modern Vision Transformer and the regression capabilities of dense neural network layers. We then apply the Laplace Approximation for probabilistic inference and uncertainty quantification (Figure \ref{fig:my_svg}). This model represents a significant methodological advance over the standard Convolutional Neural Network (CNN) architectures previously used in similar cosmological analyses (e.g., \cite{2021PASP..133d4001B}).

\subsection{Swin Transformers for Feature Extraction} \label{subsec:network}

The first and most critical task of our network is to extract meaningful features from the input kSZ maps. While CNNs have been the standard for image-based tasks in cosmology \citep{2015Natur.521..436L}, they have an inherent limitation: their convolutional kernels operate locally, making it challenging for the network to model long-range dependencies across an image efficiently. The morphology of reionization, as traced by the kSZ effect, involves correlations across a wide range of angular scales, from small ionized bubbles to large-scale patterns. Capturing these global features is essential for accurately constraining $\tau$.

For this reason, we depart from the traditional CNN approach and instead employ a Swin Transformer \citep{2021arXiv210314030L} as our primary feature extractor. The Swin (Shifted Window) Transformer is a state-of-the-art architecture that addresses key challenges in adapting Transformers from language to vision. Unlike the original Vision Transformer (ViT) \citep{2020arXiv201011929D}, which produces feature maps of a single, low resolution and has a computational complexity that is quadratic with respect to image size, the Swin Transformer is designed specifically for efficiency and multi-scale analysis.

\begin{figure}[h!]
 \centering
 \includegraphics[width=\textwidth]{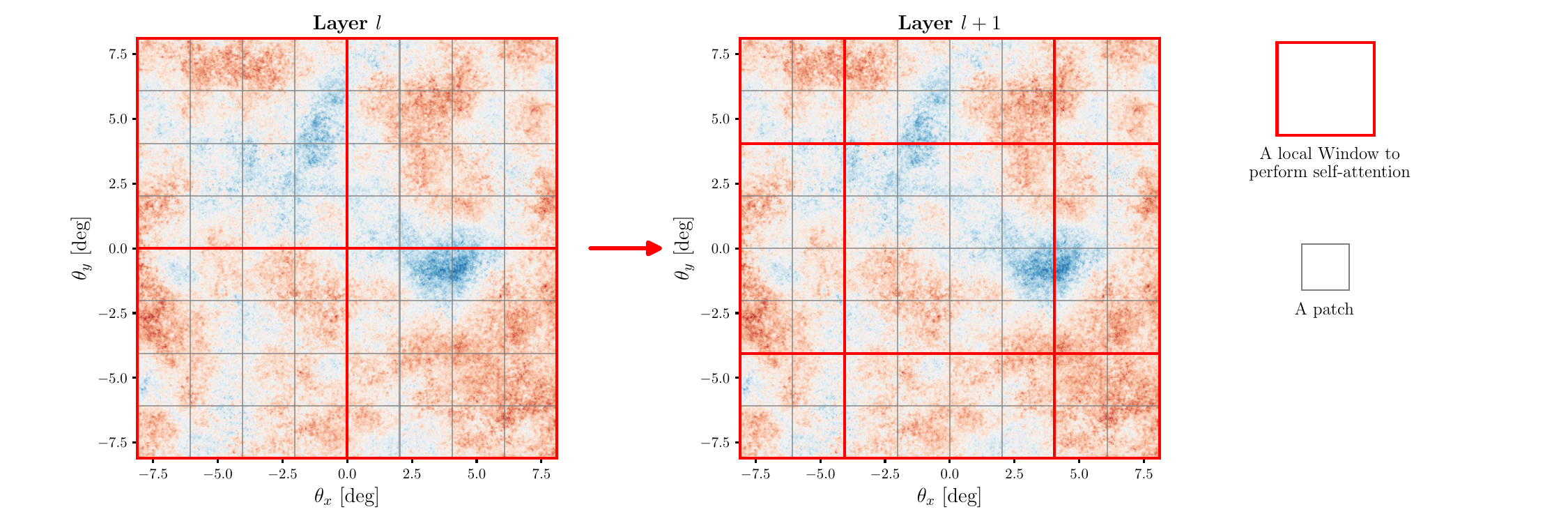}
 \caption{Illustration of the shifted-window self-attention mechanism, the core component of the Swin Transformer, shown here applied to a simulated kSZ map from our data set. In a given layer $l$ (left), the map is partitioned into regular, non-overlapping windows, and self-attention is computed only among the patches within each window. In the subsequent layer $l+1$ (right), the window grid is shifted before partitioning. This new configuration forces the self-attention calculation to cross the boundaries of the previous windows, allowing for information and features to be exchanged between them. This process is the key innovation that enables the model to learn features at multiple spatial scales in our kSZ data. Figure adapted from \citep{2021arXiv210314030L}.}
 \label{fig:shifted_window}
\end{figure}

Its architecture is built on two core principles:
\begin{itemize}
    \item Hierarchical Feature Maps: The network begins by splitting the input image into small, non-overlapping patches, treating each as a token (Figure \ref{fig:my_svg}). As these tokens pass through successive stages of the network, groups of neighboring patches are merged, effectively downsampling the spatial resolution while increasing the feature dimension. This process creates a hierarchical representation with feature maps at multiple scales, analogous to the feature pyramids common in CNNs. This design is crucial for our work as it allows the model to leverage advanced techniques for dense prediction and capture the multi-scale nature of the kSZ signal.
    \item Shifted Window Self-Attention: To maintain computational efficiency, self-attention is not calculated globally across all patches. Instead, it is computed locally within non-overlapping windows. To enable cross-window connections, which are vital for modeling global features, the window partitioning is shifted between consecutive layers. This shifted windowing scheme allows information to propagate across the entire map while ensuring that the computational complexity remains linear with respect to the input image size (Figure \ref{fig:shifted_window}). This makes the Swin Transformer a scalable and highly effective backbone for processing the high-resolution maps common in cosmology.
\end{itemize}

Furthermore, we utilize a Swin Transformer model that has been pre-trained on ImageNet, a large dataset of natural images. Through this transfer learning approach, we initialize our network with a powerful set of general-purpose feature detectors. We then fine-tune this model on our kSZ simulations, allowing it to adapt its learned features to the specific patterns and statistics of our cosmological data. This strategy significantly accelerates the training process and often leads to better performance than training a model from scratch.

The output of the Swin Transformer backbone is a high-dimensional feature vector that encodes the essential information from the input kSZ map. This vector is then passed to a sequence of fully-connected dense layers. These layers act as the regression ``head'' of our network, taking the complex features extracted by the transformer and mapping them non-linearly to a lower-dimensional space suitable for predicting a single scalar value, $\tau$.

\subsection{Uncertainty Quantification with the Laplace Approximation}
\label{sec:laplace}

A critical component of any scientific measurement is a robust estimate of its uncertainty. To move beyond the single point estimate provided by a standard trained network, we employ the Laplace Approximation (LA), a classic and efficient method for obtaining a posterior distribution over the neural network's weights \citep{2021arXiv210614806D}. The core idea of the LA is to approximate the posterior distribution, $p(\theta|\mathcal{D})$, with a Gaussian. From a Bayesian perspective, training a neural network with weight decay is equivalent to finding the maximum a posteriori (MAP) estimate, $\theta_{\text{MAP}}$, of the weights under a Gaussian prior. The posterior is proportional to the exponential of the negative loss function, $\mathcal{L}$:
\begin{equation}
    p(\theta|\mathcal{D}) \propto \exp(-\mathcal{L}(\mathcal{D};\theta)).
\end{equation}
The LA constructs a local Gaussian approximation to this posterior centered at $\theta_{\text{MAP}}$ by performing a second-order Taylor expansion of the log-posterior around this point. This yields the final Gaussian approximation for the posterior:
\begin{equation}
    p(\theta|\mathcal{D}) \approx \mathcal{N}(\theta; \theta_{\text{MAP}}, \mathbf{H}^{-1}),
    \label{eq:laplace_posterior}
\end{equation}
where the covariance matrix is the inverse of the Hessian, $\mathbf{H} = \nabla_{\theta}^{2}\mathcal{L}(\mathcal{D};\theta)|_{\theta_{\text{MAP}}}$, which captures the curvature of the loss landscape. In this work, we investigate two primary modes of applying the LA.

\subsubsection{Post-hoc Laplace Approximation}

The most direct application of the LA is the post-hoc approach. Here, we first train our combined network (Swin Transformer and the dense layers) to convergence using standard methods to find the optimal weights $\theta_{\text{MAP}}$. After training is complete, we compute the Hessian (or a scalable approximation like KFAC) at this fixed point to define the covariance of the Gaussian posterior from Equation \ref{eq:laplace_posterior}. This approach is computationally efficient and allows us to directly confer a Bayesian interpretation onto powerful, pre-trained models with minimal overhead.

\subsubsection{Online Laplace Approximation}

An alternative is the online approach, which constitutes a more fully Bayesian treatment of hyperparameter optimization. Instead of training to a MAP point and then applying the LA, the model's weights and hyperparameters (like the prior precision) are optimized \textit{jointly} by maximizing the LA to the log marginal likelihood (evidence). The log marginal likelihood, $Z$, can be approximated as:
\begin{equation}
    \log Z \approx \mathcal{L}(\mathcal{D};\theta_{\text{MAP}}) - \frac{1}{2}\log\det(\mathbf{H}) + \frac{D}{2}\log(2\pi),
    \label{eq:evidence}
\end{equation}
where $D$ is the number of parameters. By using this differentiable quantity as our optimization objective, the network learns to find a solution that not only fits the data well (low loss $\mathcal{L}$) but also occupies a wide basin in the loss landscape (low determinant of the Hessian), which naturally balances model fit and complexity \citep{2021arXiv210614806D}. Comparing these two methods allows us to determine the most robust approach for our specific cosmological problem.

\subsection{Model Hyperparameter Optimization} \label{subsec:hyperparam}
Since the Laplace Approximation has two primary variants, post-hoc and online, we implement both to compare their performance on our cosmological inference task. The optimization process for each is distinct. The post-hoc method requires first finding an optimal point-estimated model before applying the LA, while the online method integrates hyperparameter optimization directly into the training process.

\subsubsection{Post-hoc MAP Model Optimization} 
To ensure that our subsequent uncertainty analysis with the post-hoc Laplace Approximation is performed on the most accurate and well-regularized base model, we first conduct a systematic hyperparameter search to find the optimal maximum a posteriori (MAP) network configuration. We employ a large-scale, distributed hyperparameter search managed by the Ray Tune library \citep{2018arXiv180705118L}. To efficiently explore the multi-dimensional parameter space, we utilize Latin Hypercube Sampling (LHS), which provides a more uniform sampling than grid or random search, ensuring a wide range of configurations is evaluated with a limited number of trials.

Table \ref{tab:hyperparams_posthoc} summarizes the parameters used in this optimization process. The top half of the table lists the hyperparameters that were allowed to vary during our search. In addition to standard training parameters, we also varied parts of the model architecture itself, specifically the number of unfrozen blocks in the Swin Transformer backbone and the dimensions of the dense regression head. The bottom half of the table shows the auxiliary parameters that were held fixed throughout the optimization.

\begin{table}[ht]
\centering
\caption{A Summary of the Parameters used in the Post-hoc MAP Model Optimization.}
\label{tab:hyperparams_posthoc}
\begin{tabular}{ll}
\hline
\textbf{Parameter} & \textbf{Values} \\
\hline
\multicolumn{2}{c}{\textit{Varied Hyperparameters (Tuned via LHS)}} \\
\hline
Learning Rate & Log-uniform between [$10^{-6}$, $10^{-4}$] \\
Weight Decay & Log-uniform between [$10^{-5}$, $10^{-1}$] \\
Dropout Rate & Uniform between [0.1, 0.5] \\
Regression Head Dims & Integers between [16, 256] for each layer \\
Num. Unfrozen Blocks & Choice of [1, 2] \\
\hline
\multicolumn{2}{c}{\textit{Fixed Parameters}} \\
\hline
Number of Epochs & 500 \\
Optimizer & Adam \\
Loss Function & Mean Squared Error \\
LR Scheduler & ReduceLROnPlateau \\
Activation Function (Head) & ReLU \\
Batch Size & 32 \\
Early Stopping Patience & 15 epochs \\
\hline
\end{tabular}
\end{table}

For each sampled configuration, the model was trained using an early stopping criterion based on the validation loss to prevent overfitting. The hyperparameter set that yielded the lowest final validation loss was selected as our optimal MAP model for the subsequent post-hoc Laplace analysis.

\subsubsection{Online Hyperparameter Optimization} For the online Laplace variant, the optimization of both network and Laplace hyperparameters is performed jointly. We again use Ray Tune with Latin Hypercube Sampling (LHS) to manage a distributed search over the combined parameter space. In this approach, we utilize the \texttt{marglik\_training} function from the \texttt{laplace-torch} library, which trains the network by directly maximizing the log marginal likelihood from Equation \ref{eq:evidence}. This single objective function allows the training process to simultaneously learn the network weights and tune the LA's hyperparameters.

Table \ref{tab:hyperparams_online} summarizes the parameters for the online optimization. The top half lists the hyperparameters that were varied during the LHS search, including the learning rates for both the network weights and the Laplace hyperparameters. The bottom half shows the parameters that were held constant.

\begin{table}[ht]
\centering
\caption{A Summary of the Parameters used in the Online Model Optimization.}
\label{tab:hyperparams_online}
\begin{tabular}{ll}
\hline
\textbf{Parameter} & \textbf{Values} \\
\hline
\multicolumn{2}{c}{\textit{Varied Hyperparameters (Tuned via LHS)}} \\
\hline
Network Learning Rate (lr) & Log-uniform between [$10^{-5}$, $5 \times 10^{-4}$] \\
Hyperparam. Learning Rate (lr\_hyp) & Log-uniform between [$10^{-3}$, $5 \times 10^{-2}$] \\
Dropout Rate & Uniform between [0.1, 0.5] \\
Regression Head Dims & Integers between [16, 64] for each layer \\
\hline
\multicolumn{2}{c}{\textit{Fixed Parameters}} \\
\hline
Optimization Objective & Log Marginal Likelihood \\
Optimizer (Weights) & Adam \\
Hessian Structure & Full \\
Hessian Backend & AsdlGGN \\
Activation Function (Head) & ReLU \\
Batch Size & 16 \\
Burn-in Epochs & 10 \\
\hline
\end{tabular}
\end{table}

The best-performing model from the search is selected as the one that achieves the highest log marginal likelihood on the training data. This metric inherently balances data fit and model complexity, thus removing the need for a separate validation set for hyperparameter tuning, as is required in the post-hoc approach.

\section{results} \label{sec:results}
After conducting the hyperparameter optimization for both the post-hoc and online Laplace configurations, we selected the best-performing model from each approach. The post-hoc model was chosen based on the lowest validation loss, while the online model was selected based on the highest log marginal likelihood. In this section, we compare the performance of these two final models on the held-out test set to evaluate their predictive accuracy and the quality of their uncertainty estimates.

\subsection{\textbf{Predictive Performance and Uncertainty Calibration}}
\label{sec:performance_metrics}
To provide a comprehensive comparison, we evaluate the models using a suite of standard regression metrics as well as a goodness-of-fit statistic to assess the calibration of the predicted uncertainties.

\begin{itemize}
    \item \textbf{Mean Absolute Error (MAE):} The average of the absolute differences between the predicted and true values of $\tau$. It provides a straightforward measure of the typical prediction error magnitude.
    \item \textbf{Root Mean Squared Error (RMSE):} The square root of the average of the squared differences between predicted and true values. Compared to MAE, RMSE penalizes larger errors more significantly.
    \item \textbf{$R^2$ Score:} The coefficient of determination, which indicates the proportion of the variance in the true $\tau$ values that is predictable from the model's predictions. An $R^2$ score of 1 represents a perfect fit.
    \item \textbf{Pearson Correlation Coefficient (r):} Measures the linear relationship between the predicted and true values, ranging from -1 (perfect negative correlation) to +1 (perfect positive correlation).
    \item \textbf{Chi-Squared ($\chi^2$):} A goodness-of-fit metric used to evaluate the quality of the uncertainty estimates. It is calculated as the sum of the squared standardized residuals: $\chi^2 = \sum_{i} (y_i - \hat{y}_i)^2 / \sigma_i^2$, where $y_i$ is the true value, $\hat{y}_i$ is the predicted mean, and $\sigma_i^2$ is the predicted variance. This statistic measures the total deviation of the data from the model's predictions, weighted by the model's own uncertainty. For a well-calibrated model, we expect the $\chi^2$ value to be approximately equal to the number of data points in the test set ($N_{\text{test}}$).

\end{itemize}

The quantitative results for both model configurations are summarized in Table \ref{tab:results_comparison}.

\begin{table}[ht]
\centering
\caption{Performance comparison of the best post-hoc and online Laplace models on the test set.}
\label{tab:results_comparison}
\begin{tabular}{lcc}
\hline
\textbf{Metric} & \textbf{Post-hoc LA} & \textbf{Online LA} \\
\hline
MAE $\downarrow$            & \textbf{0.0012}         & 0.0017       \\
RMSE $\downarrow$           & \textbf{0.0015}         & 0.0021       \\
R² Score $\uparrow$       & \textbf{0.93}         & 0.86       \\
Pearson r $\uparrow$      & \textbf{0.96}         & 0.93       \\
Chi-Squared ($\chi^2$) & \textbf{59.27}         & 42.45       \\
\hline
\end{tabular}
\end{table}

\begin{figure*}[ht] % Use figure* for a full-width figure
    \gridline{\fig{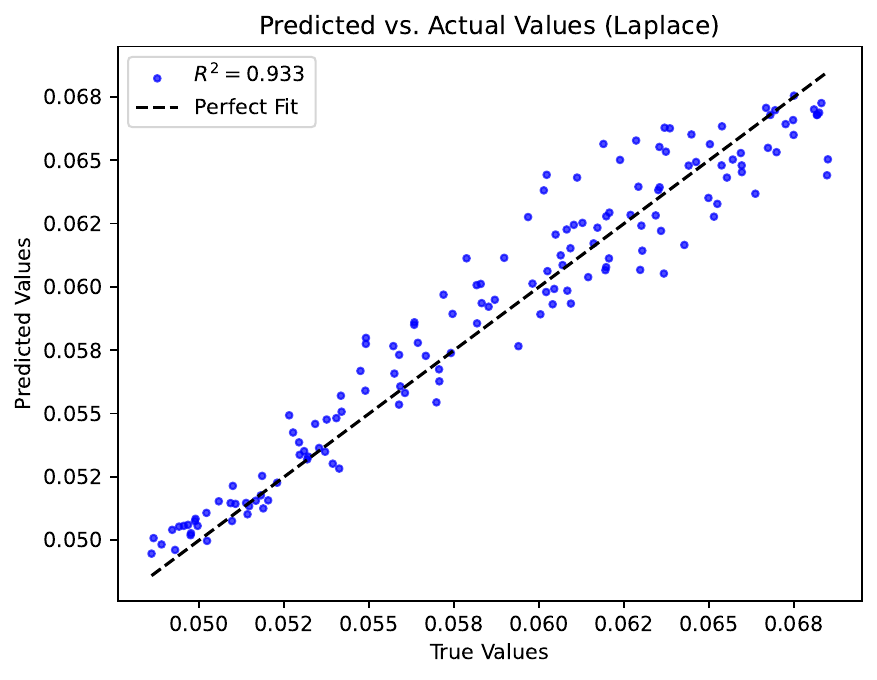}{0.48\textwidth}{(a)}
              \fig{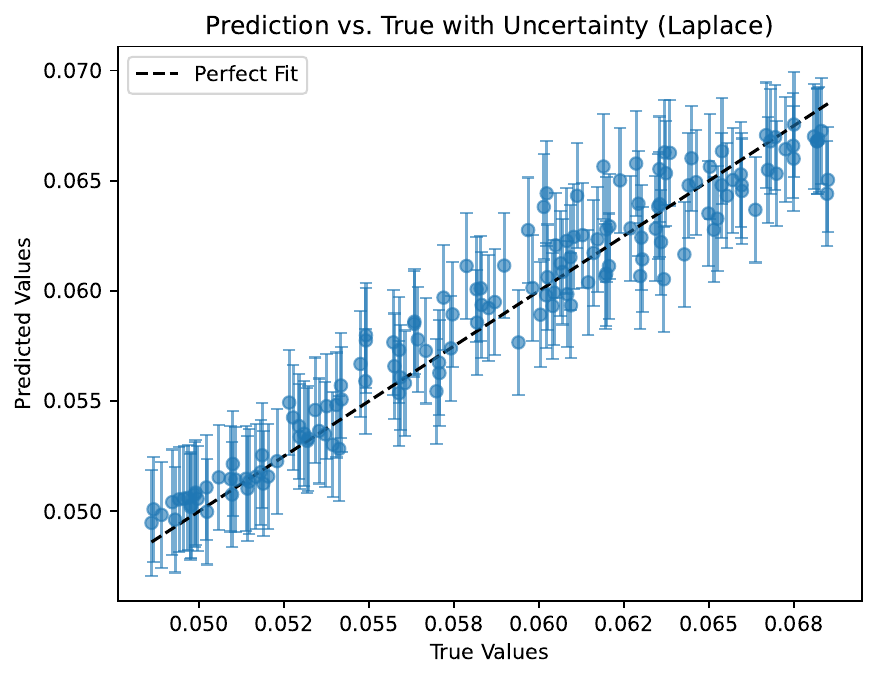}{0.48\textwidth}{(b)}}
    \gridline{\fig{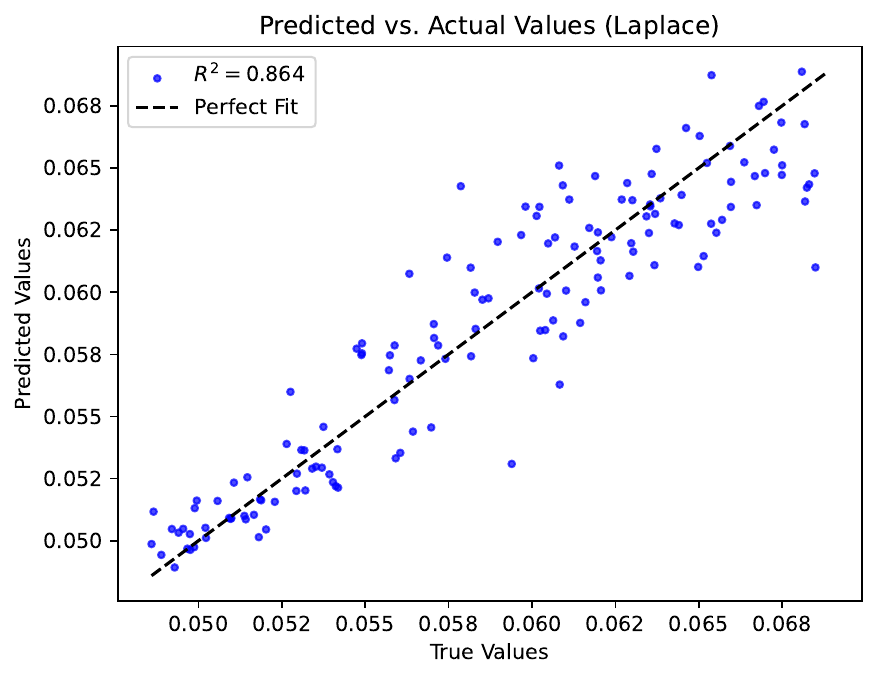}{0.48\textwidth}{(c)}
              \fig{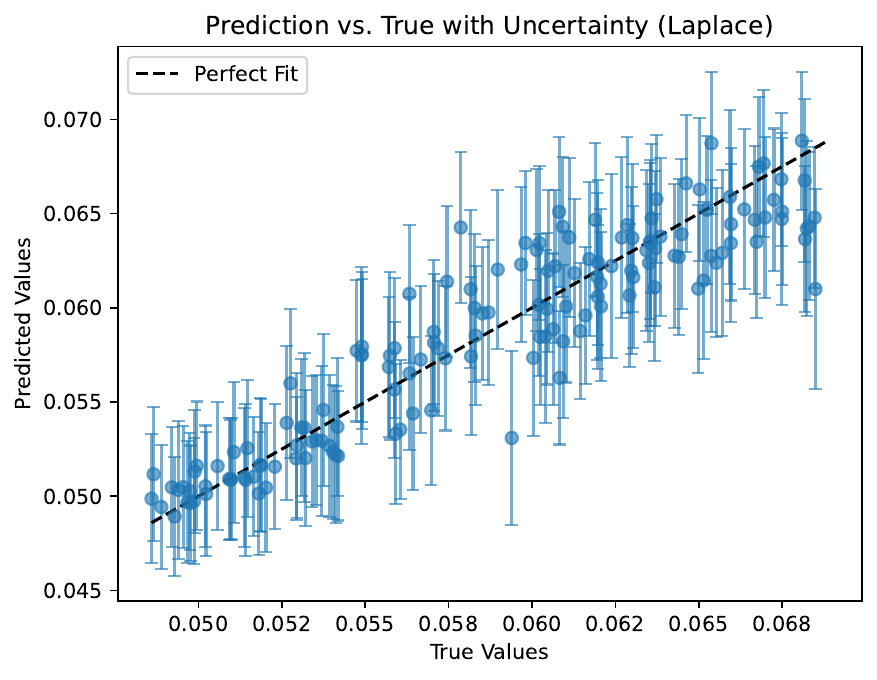}{0.48\textwidth}{(d)}}
    \caption{Visual comparison of model performance on the test set. The top row (a, b) shows results from the post-hoc Laplace model, while the bottom row (c, d) shows results for the online Laplace model. The left column (a, c) displays scatter plots of predicted vs. true $\tau$, and the right column (b, d) includes one-sigma predictive error bars.
    \label{fig:results_plots}}
\end{figure*}

To visually assess the performance and uncertainty calibration, we present scatter plots of the predicted versus true values for $\tau$ in Figure \ref{fig:results_plots}. The top row displays the results for the post-hoc Laplace model, while the bottom row shows the results for the online Laplace model. The left column shows the direct correlation between predictions and true values, while the right column includes the predicted one-sigma error bars on each point. These plots allow for a direct qualitative comparison of the accuracy and the reliability of the uncertainty estimates from each method.

\begin{figure}[ht]
\centering
\includegraphics[width=0.6\textwidth]{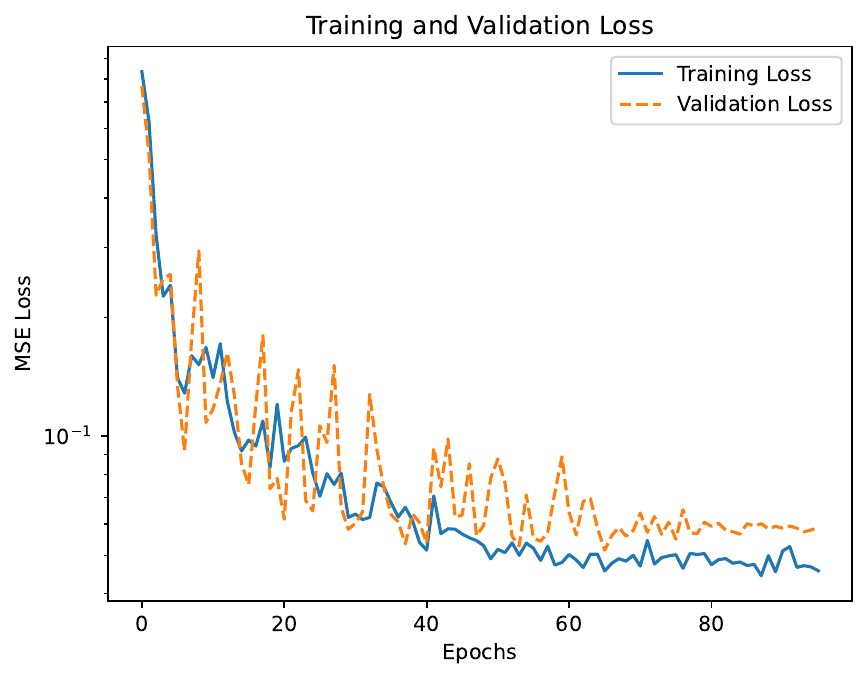}
\caption{Training and validation loss (Mean Squared Error) for the best-performing post-hoc model as a function of training epoch. The early stopping mechanism halted the training when the validation loss no longer improved, preventing overfitting.}
\label{fig:loss_curve}
\end{figure}

Since the post-hoc version demonstrates superior performance in our setup, we first examine its training dynamics. Figure \ref{fig:loss_curve} shows the training and validation loss curves for the best-performing post-hoc model. The validation loss decreases steadily before plateauing, at which point our early stopping criterion (with a patience of 15 epochs) halts the training. This behavior is a clear indication that the early stopping mechanism was effective in preventing the model from overfitting to the training data, thereby ensuring good generalization to unseen data.

\subsection{Physical Interpretability and Feature Analysis}
\label{sec:Interpretability}
To verify that the Swin Transformer exploits genuine physical structures rather than memorizing statistical artifacts, we analyzed the model's pixel-level attention using Integrated Gradients \cite{sundararajan2017axiomatic} enhanced with a Noise Tunnel \cite{smilkov2017smoothgrad}. This method generates saliency maps by integrating gradients along a path from a baseline zero-intensity image to the actual input, averaged over 50 noisy realizations to suppress high-frequency artifacts.

\begin{figure}[t!]
    \centering
    % --- Top Row: Low Tau Example ---
    \includegraphics[width=0.95\linewidth]{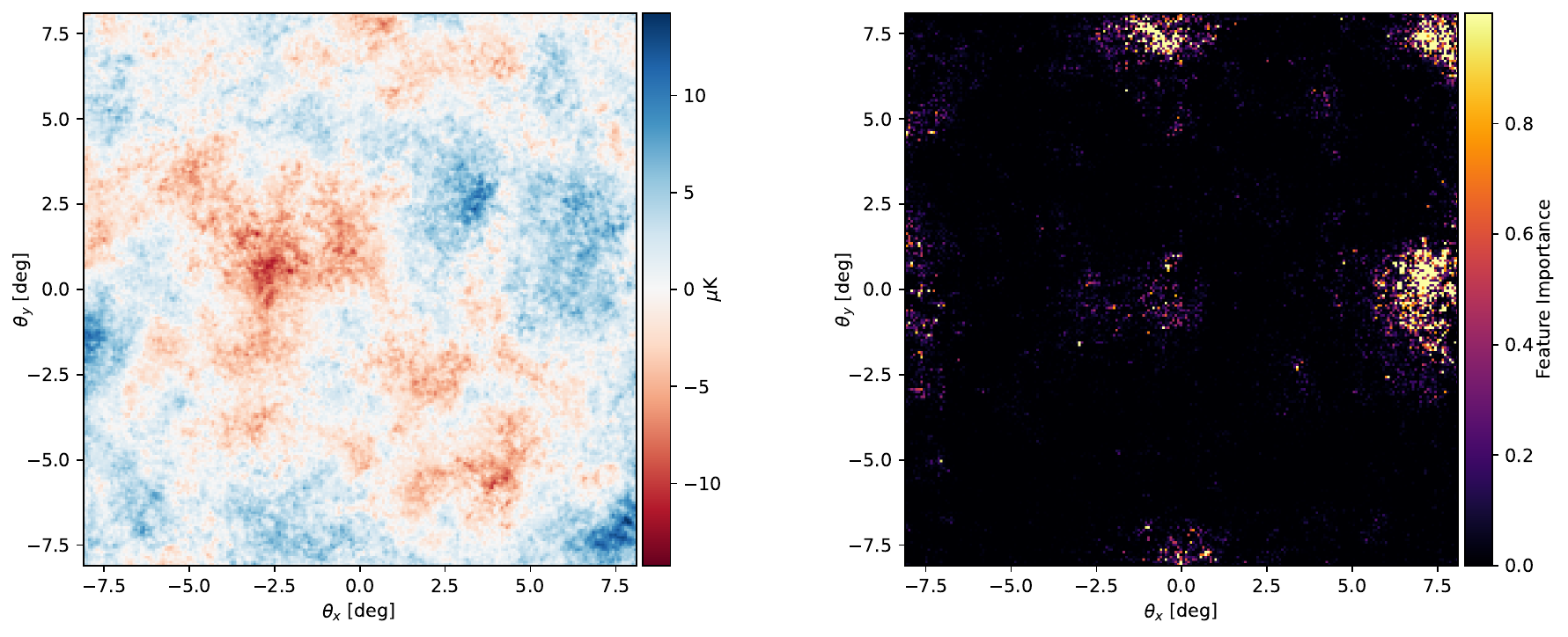} 
    
    \vspace{0.1cm} % Small gap between the two rows
    
    % --- Bottom Row: High Tau Example ---
    \includegraphics[width=0.95\linewidth]{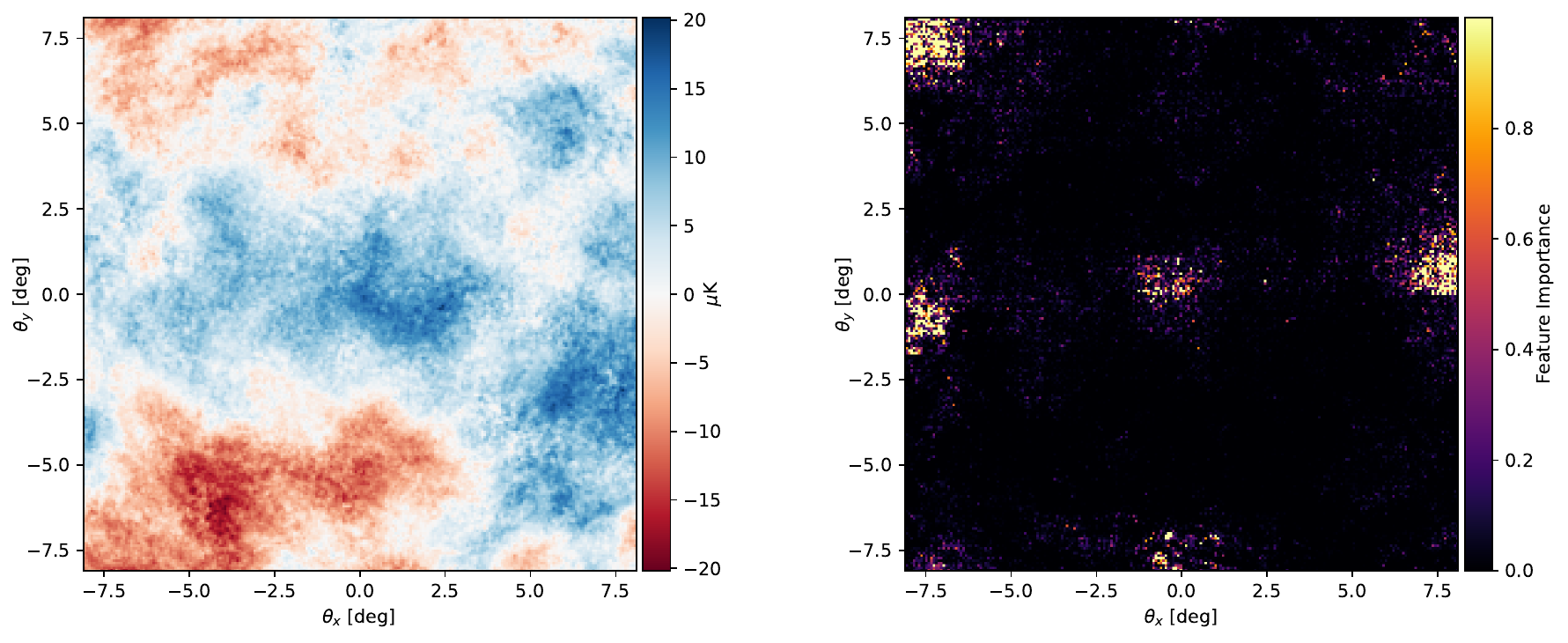}
    
    \caption{Interpretability analysis using Integrated Gradients with Noise Tunneling. 
    We display two test-set examples to visualize the features learned by the Swin Transformer. 
    \textbf{Top Row:} A sample with True $\tau = 0.04967$ (Predicted $\tau = 0.05042$). 
    \textbf{Bottom Row:} A sample with True $\tau = 0.06681$ (Predicted $\tau = 0.06658$). 
    \textbf{Left Panels:} The input kSZ maps ($\Delta T / \mu K$). 
    \textbf{Right Panels:} The corresponding saliency maps (attribution scores). 
    The network consistently assigns high attribution to the ionization fronts (the high-contrast boundaries between neutral and ionized regions) and to areas of strong velocity coherence. This confirms that the model relies on physical morphological structures, specifically the size and distribution of ionized bubbles, rather than background statistical noise to infer $\tau$.}
    \label{fig:saliency_maps}
\end{figure}

Figure \ref{fig:saliency_maps} presents the resulting saliency maps for two representative samples from the test set. In both the low-$\tau$ ($\tau \approx 0.049$) and high-$\tau$ ($\tau \approx 0.067$) regimes, the saliency maps demonstrate a strong correlation with the physical features of the kSZ signal:

\begin{enumerate}
    \item Ionization Fronts: The highest attribution scores (brightest regions in the saliency maps) consistently align with the sharp gradients at the boundaries of ionized bubbles. Physically, these fronts represent the transition between neutral and ionized hydrogen, where the contrast in electron density contributes most significantly to the integrated optical depth.
    \item Velocity Coherence: The network highlights extended, coherent structures rather than isolated pixels. This suggests the model is sensitive to the bulk velocity fields of the ionized gas, which imprint a specific morphological signature on the kSZ map (the ``Doppler boosting'' effect).
\end{enumerate}

The fact that the model tracks these morphological features (specifically the edges and sizes of the bubbles) confirms that it has learned a physical mapping between the reionization topology and the optical depth, satisfying the requirement for physical interpretability.

% --- Start of the aastex 2x2 Figure Code ---
% --- End of the aastex 2x2 Figure Code ---

\section{Conclusion and Discussion} \label{sec:discuss}
In this work, we have presented a novel machine learning framework for inferring the CMB optical depth to reionization, $\tau$, from simulated kinetic Sunyaev-Zel'dovich (kSZ) maps. Our approach utilizes a pre-trained Swin Transformer to extract the complex, non-Gaussian features from the maps and employs the Laplace Approximation (LA) to provide principled, data-driven uncertainty estimates for our predictions. We conducted a detailed comparison between two primary modes of applying the LA: a post-hoc method applied to a pre-trained MAP model, and an online method where network and LA hyperparameters are optimized jointly.

Our results, summarized in Figure \ref{fig:results_plots}, demonstrate the success of this approach and show a clear performance difference between the two LA configurations. The post-hoc model achieved a high degree of accuracy, with a coefficient of determination of $R^2 = 0.933$. The scatter plot in Figure \ref{fig:results_plots}(a) shows a tight linear correlation between the predicted and true values of $\tau$, indicating that the model is a reliable estimator. Furthermore, the error bar plot in Figure \ref{fig:results_plots}(b) shows that the model's predictions are consistent with the true values within their one-sigma uncertainties, suggesting the LA provides well-calibrated error estimates.

In contrast, the online model, while still showing a significant correlation, performed less accurately, achieving an $R^2 = 0.866$. The increased scatter in Figure \ref{fig:results_plots}(c) and (d) indicates lower predictive power compared to the post-hoc approach. We attribute this performance difference to the optimization strategy. The post-hoc method benefits from being applied to a model whose weights have already been optimized to a high-performing MAP solution via an extensive, dedicated hyperparameter search. The online method's joint optimization of weights and hyperparameters, while more fully Bayesian, appears to have settled in a less optimal region of the parameter space for the network weights. This suggests that for problems where a high-quality point-estimated model can be found, the post-hoc LA is the more effective and straightforward approach for adding robust uncertainty quantification.

While this proof-of-concept work is based on idealized simulations, it demonstrates the significant potential of combining modern deep learning architectures with efficient Bayesian approximation techniques for cosmological inference. The ability to extract $\tau$ from kSZ maps offers a powerful and independent complement to constraints from the 21 cm signal and direct CMB measurements. Future work will involve applying this framework to more realistic simulations that include instrumental noise and astrophysical foregrounds, which will be a critical step toward applying these techniques to data from upcoming CMB surveys like the Simons Observatory and CMB-S4. As discussed above in Sec.~\ref{sec:ksz_maps}, one possible approach could include applying a Wiener filter to CMB maps to extract the reionization-era signal. However, such an approach must contend with potential limitations, including the late-time kSZ and the effect of foregrounds. We plan to investigate such an approach in future work.

In conclusion, we have shown that a Swin Transformer combined with a post-hoc Laplace Approximation is a \textbf{potentially} powerful and computationally efficient tool for constraining the optical depth to reionization from kSZ maps. This method not only provides accurate point estimates but also the principled uncertainty quantification that is essential for robust scientific analysis.

\begin{acknowledgments}
We thank James Aguirre for insightful discussions about this work. FFK and PL are supported by Simons Foundation award number 00007127. AKS and PL are supported by the U. S. National Science Foundation grant \#2206602. This work used Bridges-2 at the Pittsburgh Computing Center through allocation AST180004 from the Advanced Cyberinfrastructure Coordination Ecosystem: Services \& Support (ACCESS) program, which is supported by U. S. National Science Foundation grants \#2138259, \#2138286, \#2138307, \#2137603, and \#2138296 \citep{10.1145/3569951.3597559}. This work used the RebelX cluster at the University of Nevada, Las Vegas, which is supported by the U. S. National Science Foundation grant \#2117941.
\end{acknowledgments}

\software{NumPy \citep{harris2020array}, Matplotlib \citep{Hunter:2007}, Astropy \citep{astropy:2013, astropy:2018, astropy:2022}, PyTorch \citep{2019arXiv191201703P}.}


\bibliography{PASPsample631}{}

@ARTICLE{2001PhR...349..125B,
       author = {{Barkana}, R. and {Loeb}, A.},
        title = "{In the beginning: the first sources of light and the reionization of the universe}",
      journal = {\physrep},
         year = 2001,
        month = jul,
       volume = {349},
       number = {2},
        pages = {125-238},
          doi = {10.1016/S0370-1573(01)00019-9},
archivePrefix = {arXiv},
       eprint = {astro-ph/0010468},
 primaryClass = {astro-ph},
       adsurl = {https://ui.adsabs.harvard.edu/abs/2001PhR...349..125B}
}

@ARTICLE{2019JCAP...02..056A,
       author = {{Ade}, Peter and {Aguirre}, James and {Ahmed}, Zeeshan and {Aiola}, Simone and {Ali}, Aamir and {Alonso}, David and {Alvarez}, Marcelo A. and {Arnold}, Kam and {Ashton}, Peter and {Austermann}, Jason and {Awan}, Humna and {Baccigalupi}, Carlo and {Baildon}, Taylor and {Barron}, Darcy and {Battaglia}, Nick and {Battye}, Richard and {Baxter}, Eric and {Bazarko}, Andrew and {Beall}, James A. and {Bean}, Rachel and {Beck}, Dominic and {Beckman}, Shawn and {Beringue}, Benjamin and {Bianchini}, Federico and {Boada}, Steven and {Boettger}, David and {Bond}, J. Richard and {Borrill}, Julian and {Brown}, Michael L. and {Bruno}, Sarah Marie and {Bryan}, Sean and {Calabrese}, Erminia and {Calafut}, Victoria and {Calisse}, Paolo and {Carron}, Julien and {Challinor}, Anthony and {Chesmore}, Grace and {Chinone}, Yuji and {Chluba}, Jens and {Cho}, Hsiao-Mei Sherry and {Choi}, Steve and {Coppi}, Gabriele and {Cothard}, Nicholas F. and {Coughlin}, Kevin and {Crichton}, Devin and {Crowley}, Kevin D. and {Crowley}, Kevin T. and {Cukierman}, Ari and {D'Ewart}, John M. and {D{\"u}nner}, Rolando and {de Haan}, Tijmen and {Devlin}, Mark and {Dicker}, Simon and {Didier}, Joy and {Dobbs}, Matt and {Dober}, Bradley and {Duell}, Cody J. and {Duff}, Shannon and {Duivenvoorden}, Adri and {Dunkley}, Jo and {Dusatko}, John and {Errard}, Josquin and {Fabbian}, Giulio and {Feeney}, Stephen and {Ferraro}, Simone and {Flux{\`a}}, Pedro and {Freese}, Katherine and {Frisch}, Josef C. and {Frolov}, Andrei and {Fuller}, George and {Fuzia}, Brittany and {Galitzki}, Nicholas and {Gallardo}, Patricio A. and {Tomas Galvez Ghersi}, Jose and {Gao}, Jiansong and {Gawiser}, Eric and {Gerbino}, Martina and {Gluscevic}, Vera and {Goeckner-Wald}, Neil and {Golec}, Joseph and {Gordon}, Sam and {Gralla}, Megan and {Green}, Daniel and {Grigorian}, Arpi and {Groh}, John and {Groppi}, Chris and {Guan}, Yilun and {Gudmundsson}, Jon E. and {Han}, Dongwon and {Hargrave}, Peter and {Hasegawa}, Masaya and {Hasselfield}, Matthew and {Hattori}, Makoto and {Haynes}, Victor and {Hazumi}, Masashi and {He}, Yizhou and {Healy}, Erin and {Henderson}, Shawn W. and {Hervias-Caimapo}, Carlos and {Hill}, Charles A. and {Hill}, J. Colin and {Hilton}, Gene and {Hilton}, Matt and {Hincks}, Adam D. and {Hinshaw}, Gary and {Hlo{\v{z}}ek}, Ren{\'e}e and {Ho}, Shirley and {Ho}, Shuay-Pwu Patty and {Howe}, Logan and {Huang}, Zhiqi and {Hubmayr}, Johannes and {Huffenberger}, Kevin and {Hughes}, John P. and {Ijjas}, Anna and {Ikape}, Margaret and {Irwin}, Kent and {Jaffe}, Andrew H. and {Jain}, Bhuvnesh and {Jeong}, Oliver and {Kaneko}, Daisuke and {Karpel}, Ethan D. and {Katayama}, Nobuhiko and {Keating}, Brian and {Kernasovskiy}, Sarah S. and {Keskitalo}, Reijo and {Kisner}, Theodore and {Kiuchi}, Kenji and {Klein}, Jeff and {Knowles}, Kenda and {Koopman}, Brian and {Kosowsky}, Arthur and {Krachmalnicoff}, Nicoletta and {Kuenstner}, Stephen E. and {Kuo}, Chao-Lin and {Kusaka}, Akito and {Lashner}, Jacob and {Lee}, Adrian and {Lee}, Eunseong and {Leon}, David and {Leung}, Jason S. -Y. and {Lewis}, Antony and {Li}, Yaqiong and {Li}, Zack and {Limon}, Michele and {Linder}, Eric and {Lopez-Caraballo}, Carlos and {Louis}, Thibaut and {Lowry}, Lindsay and {Lungu}, Marius and {Madhavacheril}, Mathew and {Mak}, Daisy and {Maldonado}, Felipe and {Mani}, Hamdi and {Mates}, Ben and {Matsuda}, Frederick and {Maurin}, Lo{\"\i}c and {Mauskopf}, Phil and {May}, Andrew and {McCallum}, Nialh and {McKenney}, Chris and {McMahon}, Jeff and {Meerburg}, P. Daniel and {Meyers}, Joel and {Miller}, Amber and {Mirmelstein}, Mark and {Moodley}, Kavilan and {Munchmeyer}, Moritz and {Munson}, Charles and {Naess}, Sigurd and {Nati}, Federico and {Navaroli}, Martin and {Newburgh}, Laura and {Nguyen}, Ho Nam and {Niemack}, Michael and {Nishino}, Haruki and {Orlowski-Scherer}, John and {Page}, Lyman and {Partridge}, Bruce and {Peloton}, Julien and {Perrotta}, Francesca and {Piccirillo}, Lucio and {Pisano}, Giampaolo and {Poletti}, Davide and {Puddu}, Roberto and {Puglisi}, Giuseppe and {Raum}, Chris and {Reichardt}, Christian L. and {Remazeilles}, Mathieu and {Rephaeli}, Yoel and {Riechers}, Dominik and {Rojas}, Felipe and {Roy}, Anirban and {Sadeh}, Sharon and {Sakurai}, Yuki and {Salatino}, Maria and {Sathyanarayana Rao}, Mayuri and {Schaan}, Emmanuel and {Schmittfull}, Marcel and {Sehgal}, Neelima and {Seibert}, Joseph},
        title = "{The Simons Observatory: science goals and forecasts}",
      journal = {\jcap},
         year = 2019,
        month = feb,
       volume = {2019},
       number = {2},
          eid = {056},
        pages = {056},
          doi = {10.1088/1475-7516/2019/02/056},
archivePrefix = {arXiv},
       eprint = {1808.07445},
 primaryClass = {astro-ph.CO},
       adsurl = {https://ui.adsabs.harvard.edu/abs/2019JCAP...02..056A}
}

@ARTICLE{2020A&A...641A...6P,
       author = {{Planck Collaboration} and {Aghanim}, N. and {Akrami}, Y. and {Ashdown}, M. and {Aumont}, J. and {Baccigalupi}, C. and {Ballardini}, M. and {Banday}, A.~J. and {Barreiro}, R.~B. and {Bartolo}, N. and {Basak}, S. and {Battye}, R. and {Benabed}, K. and {Bernard}, J. -P. and {Bersanelli}, M. and {Bielewicz}, P. and {Bock}, J.~J. and {Bond}, J.~R. and {Borrill}, J. and {Bouchet}, F.~R. and {Boulanger}, F. and {Bucher}, M. and {Burigana}, C. and {Butler}, R.~C. and {Calabrese}, E. and {Cardoso}, J. -F. and {Carron}, J. and {Challinor}, A. and {Chiang}, H.~C. and {Chluba}, J. and {Colombo}, L.~P.~L. and {Combet}, C. and {Contreras}, D. and {Crill}, B.~P. and {Cuttaia}, F. and {de Bernardis}, P. and {de Zotti}, G. and {Delabrouille}, J. and {Delouis}, J. -M. and {Di Valentino}, E. and {Diego}, J.~M. and {Dor{\'e}}, O. and {Douspis}, M. and {Ducout}, A. and {Dupac}, X. and {Dusini}, S. and {Efstathiou}, G. and {Elsner}, F. and {En{\ss}lin}, T.~A. and {Eriksen}, H.~K. and {Fantaye}, Y. and {Farhang}, M. and {Fergusson}, J. and {Fernandez-Cobos}, R. and {Finelli}, F. and {Forastieri}, F. and {Frailis}, M. and {Fraisse}, A.~A. and {Franceschi}, E. and {Frolov}, A. and {Galeotta}, S. and {Galli}, S. and {Ganga}, K. and {G{\'e}nova-Santos}, R.~T. and {Gerbino}, M. and {Ghosh}, T. and {Gonz{\'a}lez-Nuevo}, J. and {G{\'o}rski}, K.~M. and {Gratton}, S. and {Gruppuso}, A. and {Gudmundsson}, J.~E. and {Hamann}, J. and {Handley}, W. and {Hansen}, F.~K. and {Herranz}, D. and {Hildebrandt}, S.~R. and {Hivon}, E. and {Huang}, Z. and {Jaffe}, A.~H. and {Jones}, W.~C. and {Karakci}, A. and {Keih{\"a}nen}, E. and {Keskitalo}, R. and {Kiiveri}, K. and {Kim}, J. and {Kisner}, T.~S. and {Knox}, L. and {Krachmalnicoff}, N. and {Kunz}, M. and {Kurki-Suonio}, H. and {Lagache}, G. and {Lamarre}, J. -M. and {Lasenby}, A. and {Lattanzi}, M. and {Lawrence}, C.~R. and {Le Jeune}, M. and {Lemos}, P. and {Lesgourgues}, J. and {Levrier}, F. and {Lewis}, A. and {Liguori}, M. and {Lilje}, P.~B. and {Lilley}, M. and {Lindholm}, V. and {L{\'o}pez-Caniego}, M. and {Lubin}, P.~M. and {Ma}, Y. -Z. and {Mac{\'\i}as-P{\'e}rez}, J.~F. and {Maggio}, G. and {Maino}, D. and {Mandolesi}, N. and {Mangilli}, A. and {Marcos-Caballero}, A. and {Maris}, M. and {Martin}, P.~G. and {Martinelli}, M. and {Mart{\'\i}nez-Gonz{\'a}lez}, E. and {Matarrese}, S. and {Mauri}, N. and {McEwen}, J.~D. and {Meinhold}, P.~R. and {Melchiorri}, A. and {Mennella}, A. and {Migliaccio}, M. and {Millea}, M. and {Mitra}, S. and {Miville-Desch{\^e}nes}, M. -A. and {Molinari}, D. and {Montier}, L. and {Morgante}, G. and {Moss}, A. and {Natoli}, P. and {N{\o}rgaard-Nielsen}, H.~U. and {Pagano}, L. and {Paoletti}, D. and {Partridge}, B. and {Patanchon}, G. and {Peiris}, H.~V. and {Perrotta}, F. and {Pettorino}, V. and {Piacentini}, F. and {Polastri}, L. and {Polenta}, G. and {Puget}, J. -L. and {Rachen}, J.~P. and {Reinecke}, M. and {Remazeilles}, M. and {Renzi}, A. and {Rocha}, G. and {Rosset}, C. and {Roudier}, G. and {Rubi{\~n}o-Mart{\'\i}n}, J.~A. and {Ruiz-Granados}, B. and {Salvati}, L. and {Sandri}, M. and {Savelainen}, M. and {Scott}, D. and {Shellard}, E.~P.~S. and {Sirignano}, C. and {Sirri}, G. and {Spencer}, L.~D. and {Sunyaev}, R. and {Suur-Uski}, A. -S. and {Tauber}, J.~A. and {Tavagnacco}, D. and {Tenti}, M. and {Toffolatti}, L. and {Tomasi}, M. and {Trombetti}, T. and {Valenziano}, L. and {Valiviita}, J. and {Van Tent}, B. and {Vibert}, L. and {Vielva}, P. and {Villa}, F. and {Vittorio}, N. and {Wandelt}, B.~D. and {Wehus}, I.~K. and {White}, M. and {White}, S.~D.~M. and {Zacchei}, A. and {Zonca}, A.},
        title = "{Planck 2018 results. VI. Cosmological parameters}",
      journal = {\aap},
         year = 2020,
        month = sep,
       volume = {641},
          eid = {A6},
        pages = {A6},
          doi = {10.1051/0004-6361/201833910},
archivePrefix = {arXiv},
       eprint = {1807.06209},
 primaryClass = {astro-ph.CO},
       adsurl = {https://ui.adsabs.harvard.edu/abs/2020A&A...641A...6P}
}

@ARTICLE{2020ApJ...903..104P,
       author = {{Petroff}, Matthew A. and {Addison}, Graeme E. and {Bennett}, Charles L. and {Weiland}, Janet L.},
        title = "{Full-sky Cosmic Microwave Background Foreground Cleaning Using Machine Learning}",
      journal = {\apj},
         year = 2020,
        month = nov,
       volume = {903},
       number = {2},
          eid = {104},
        pages = {104},
          doi = {10.3847/1538-4357/abb9a7},
archivePrefix = {arXiv},
       eprint = {2004.11507},
 primaryClass = {astro-ph.CO},
       adsurl = {https://ui.adsabs.harvard.edu/abs/2020ApJ...903..104P}
}

@ARTICLE{2021arXiv210614806D,
       author = {{Daxberger}, Erik and {Kristiadi}, Agustinus and {Immer}, Alexander and {Eschenhagen}, Runa and {Bauer}, Matthias and {Hennig}, Philipp},
        title = "{Laplace Redux -- Effortless Bayesian Deep Learning}",
      journal = {arXiv e-prints},
         year = 2021,
        month = jun,
          eid = {arXiv:2106.14806},
        pages = {arXiv:2106.14806},
          doi = {10.48550/arXiv.2106.14806},
archivePrefix = {arXiv},
       eprint = {2106.14806},
 primaryClass = {cs.LG},
       adsurl = {https://ui.adsabs.harvard.edu/abs/2021arXiv210614806D}
}

@ARTICLE{2021arXiv210314030L,
       author = {{Liu}, Ze and {Lin}, Yutong and {Cao}, Yue and {Hu}, Han and {Wei}, Yixuan and {Zhang}, Zheng and {Lin}, Stephen and {Guo}, Baining},
        title = "{Swin Transformer: Hierarchical Vision Transformer using Shifted Windows}",
      journal = {arXiv e-prints},
         year = 2021,
        month = mar,
          eid = {arXiv:2103.14030},
        pages = {arXiv:2103.14030},
          doi = {10.48550/arXiv.2103.14030},
archivePrefix = {arXiv},
       eprint = {2103.14030},
 primaryClass = {cs.CV},
       adsurl = {https://ui.adsabs.harvard.edu/abs/2021arXiv210314030L}
}

@ARTICLE{2020arXiv201011929D,
       author = {{Dosovitskiy}, Alexey and {Beyer}, Lucas and {Kolesnikov}, Alexander and {Weissenborn}, Dirk and {Zhai}, Xiaohua and {Unterthiner}, Thomas and {Dehghani}, Mostafa and {Minderer}, Matthias and {Heigold}, Georg and {Gelly}, Sylvain and {Uszkoreit}, Jakob and {Houlsby}, Neil},
        title = "{An Image is Worth 16x16 Words: Transformers for Image Recognition at Scale}",
      journal = {arXiv e-prints},
         year = 2020,
        month = oct,
          eid = {arXiv:2010.11929},
        pages = {arXiv:2010.11929},
          doi = {10.48550/arXiv.2010.11929},
archivePrefix = {arXiv},
       eprint = {2010.11929},
 primaryClass = {cs.CV},
       adsurl = {https://ui.adsabs.harvard.edu/abs/2020arXiv201011929D}
}

@ARTICLE{2018arXiv180705118L,
       author = {{Liaw}, Richard and {Liang}, Eric and {Nishihara}, Robert and {Moritz}, Philipp and {Gonzalez}, Joseph E. and {Stoica}, Ion},
        title = "{Tune: A Research Platform for Distributed Model Selection and Training}",
      journal = {arXiv e-prints},
         year = 2018,
        month = jul,
          eid = {arXiv:1807.05118},
        pages = {arXiv:1807.05118},
          doi = {10.48550/arXiv.1807.05118},
archivePrefix = {arXiv},
       eprint = {1807.05118},
 primaryClass = {cs.LG},
       adsurl = {https://ui.adsabs.harvard.edu/abs/2018arXiv180705118L}
}

@ARTICLE{2013ApJ...776...81B,
       author = {{Battaglia}, N. and {Trac}, H. and {Cen}, R. and {Loeb}, A.},
        title = "{Reionization on Large Scales. I. A Parametric Model Constructed from Radiation-hydrodynamic Simulations}",
      journal = {\apj},
         year = 2013,
        month = oct,
       volume = {776},
       number = {2},
          eid = {81},
        pages = {81},
          doi = {10.1088/0004-637X/776/2/81},
archivePrefix = {arXiv},
       eprint = {1211.2821},
 primaryClass = {astro-ph.CO},
       adsurl = {https://ui.adsabs.harvard.edu/abs/2013ApJ...776...81B}
}

@ARTICLE{2013ApJ...776...83B,
       author = {{Battaglia}, N. and {Natarajan}, A. and {Trac}, H. and {Cen}, R. and {Loeb}, A.},
        title = "{Reionization on Large Scales. III. Predictions for Low-l Cosmic Microwave Background Polarization and High-l Kinetic Sunyaev-Zel'dovich Observables}",
      journal = {\apj},
         year = 2013,
        month = oct,
       volume = {776},
       number = {2},
          eid = {83},
        pages = {83},
          doi = {10.1088/0004-637X/776/2/83},
archivePrefix = {arXiv},
       eprint = {1211.2832},
 primaryClass = {astro-ph.CO},
       adsurl = {https://ui.adsabs.harvard.edu/abs/2013ApJ...776...83B}
}

@ARTICLE{2020ApJ...899...40L,
       author = {{La Plante}, Paul and {Lidz}, Adam and {Aguirre}, James and {Kohn}, Saul},
        title = "{The 21 cm kSZ-kSZ Bispectrum during the Epoch of Reionization}",
      journal = {\apj},
         year = 2020,
        month = aug,
       volume = {899},
       number = {1},
          eid = {40},
        pages = {40},
          doi = {10.3847/1538-4357/aba2ed},
archivePrefix = {arXiv},
       eprint = {2005.07206},
 primaryClass = {astro-ph.CO},
       adsurl = {https://ui.adsabs.harvard.edu/abs/2020ApJ...899...40L}
}

@ARTICLE{2022ApJ...928..162L,
       author = {{La Plante}, Paul and {Sipple}, Jackson and {Lidz}, Adam},
        title = "{Prospects for kSZ$^{2}$-Galaxy Cross-correlations during Reionization}",
      journal = {\apj},
         year = 2022,
        month = apr,
       volume = {928},
       number = {2},
          eid = {162},
        pages = {162},
          doi = {10.3847/1538-4357/ac5752},
archivePrefix = {arXiv},
       eprint = {2111.13717},
 primaryClass = {astro-ph.CO},
       adsurl = {https://ui.adsabs.harvard.edu/abs/2022ApJ...928..162L}
}

@ARTICLE{2025ApJ...991..195Z,
       author = {{Zhou}, Meng and {La Plante}, Paul and {Lidz}, Adam and {Mao}, Yi and {Ma}, Yin-Zhe},
        title = "{Prospects for kSZ$^{2}$─21 cm$^{2}$ Cross Correlations during Reionization}",
      journal = {\apj},
         year = 2025,
        month = oct,
       volume = {991},
       number = {2},
          eid = {195},
        pages = {195},
          doi = {10.3847/1538-4357/adfb67},
archivePrefix = {arXiv},
       eprint = {2503.09462},
 primaryClass = {astro-ph.CO},
       adsurl = {https://ui.adsabs.harvard.edu/abs/2025ApJ...991..195Z}
}

@ARTICLE{2019ApJ...880..110L,
       author = {{La Plante}, Paul and {Ntampaka}, Michelle},
        title = "{Machine Learning Applied to the Reionization History of the Universe in the 21 cm Signal}",
      journal = {\apj},
         year = 2019,
        month = aug,
       volume = {880},
       number = {2},
          eid = {110},
        pages = {110},
          doi = {10.3847/1538-4357/ab2983},
archivePrefix = {arXiv},
       eprint = {1810.08211},
 primaryClass = {astro-ph.CO},
       adsurl = {https://ui.adsabs.harvard.edu/abs/2019ApJ...880..110L}
}

@ARTICLE{2021PASP..133d4001B,
       author = {{Billings}, Tashalee S. and {La Plante}, Paul and {Aguirre}, James E.},
        title = "{Extracting the Optical Depth to Reionization {\ensuremath{\tau}} from 21 cm Data Using Machine Learning Techniques}",
      journal = {\pasp},
         year = 2021,
        month = apr,
       volume = {133},
       number = {1022},
          eid = {044001},
        pages = {044001},
          doi = {10.1088/1538-3873/abe9a0},
       adsurl = {https://ui.adsabs.harvard.edu/abs/2021PASP..133d4001B}
}

@ARTICLE{2022PASP..134d4001Z,
       author = {{Zhou}, Yihao and {La Plante}, Paul},
        title = "{Understanding the Impact of Semi-numeric Reionization Models when Using CNNs}",
      journal = {\pasp},
         year = 2022,
        month = apr,
       volume = {134},
       number = {1034},
          eid = {044001},
        pages = {044001},
          doi = {10.1088/1538-3873/ac5f5d},
archivePrefix = {arXiv},
       eprint = {2112.03443},
 primaryClass = {astro-ph.CO},
       adsurl = {https://ui.adsabs.harvard.edu/abs/2022PASP..134d4001Z}
}

@ARTICLE{1972CoASP...4..173S,
       author = {{Sunyaev}, R.~A. and {Zeldovich}, Ya. B.},
        title = "{The Observations of Relic Radiation as a Test of the Nature of X-Ray Radiation from the Clusters of Galaxies}",
      journal = {Comments on Astrophysics and Space Physics},
         year = 1972,
        month = nov,
       volume = {4},
        pages = {173},
       adsurl = {https://ui.adsabs.harvard.edu/abs/1972CoASP...4..173S}
}

@ARTICLE{2016ApJ...824..118A,
       author = {{Alvarez}, Marcelo A.},
        title = "{The Kinetic Sunyaev-Zel{\textquoteright}dovich Effect from Reionization: Simulated Full-sky Maps at Arcminute Resolution}",
      journal = {\apj},
         year = 2016,
        month = jun,
       volume = {824},
       number = {2},
          eid = {118},
        pages = {118},
          doi = {10.3847/0004-637X/824/2/118},
archivePrefix = {arXiv},
       eprint = {1511.02846},
 primaryClass = {astro-ph.CO},
       adsurl = {https://ui.adsabs.harvard.edu/abs/2016ApJ...824..118A}
}

@ARTICLE{2017ApJ...841...87L,
       author = {{La Plante}, Paul and {Trac}, Hy and {Croft}, Rupert and {Cen}, Renyue},
        title = "{Helium Reionization Simulations. II. Signatures of Quasar Activity on the IGM}",
      journal = {\apj},
         year = 2017,
        month = jun,
       volume = {841},
       number = {2},
          eid = {87},
        pages = {87},
          doi = {10.3847/1538-4357/aa7136},
archivePrefix = {arXiv},
       eprint = {1610.02047},
 primaryClass = {astro-ph.CO},
       adsurl = {https://ui.adsabs.harvard.edu/abs/2017ApJ...841...87L}
}

@ARTICLE{2015Natur.521..436L,
       author = {{LeCun}, Yann and {Bengio}, Yoshua and {Hinton}, Geoffrey},
        title = "{Deep learning}",
      journal = {\nat},
         year = 2015,
        month = may,
       volume = {521},
       number = {7553},
        pages = {436-444},
          doi = {10.1038/nature14539},
       adsurl = {https://ui.adsabs.harvard.edu/abs/2015Natur.521..436L}
}

@BOOK{2013fgu..book.....L,
       author = {{Loeb}, Abraham and {Furlanetto}, Steven R.},
        title = "{The First Galaxies in the Universe}",
         year = 2013,
    publisher = "{Princeton University Press}",
       adsurl = {https://ui.adsabs.harvard.edu/abs/2013fgu..book.....L}
}

@ARTICLE{2006PhR...433..181F,
       author = {{Furlanetto}, Steven R. and {Oh}, S. Peng and {Briggs}, Frank H.},
        title = "{Cosmology at low frequencies: The 21 cm transition and the high-redshift Universe}",
      journal = {\physrep},
         year = 2006,
        month = oct,
       volume = {433},
       number = {4-6},
        pages = {181-301},
          doi = {10.1016/j.physrep.2006.08.002},
archivePrefix = {arXiv},
       eprint = {astro-ph/0608032},
 primaryClass = {astro-ph},
       adsurl = {https://ui.adsabs.harvard.edu/abs/2006PhR...433..181F}
}

@ARTICLE{2016PhRvD..93d3013L,
       author = {{Liu}, Adrian and {Pritchard}, Jonathan R. and {Allison}, Rupert and {Parsons}, Aaron R. and {Seljak}, Uro{\v{s}} and {Sherwin}, Blake D.},
        title = "{Eliminating the optical depth nuisance from the CMB with 21 cm cosmology}",
      journal = {\prd},
         year = 2016,
        month = feb,
       volume = {93},
       number = {4},
          eid = {043013},
        pages = {043013},
          doi = {10.1103/PhysRevD.93.043013},
archivePrefix = {arXiv},
       eprint = {1509.08463},
 primaryClass = {astro-ph.CO},
       adsurl = {https://ui.adsabs.harvard.edu/abs/2016PhRvD..93d3013L}
}

@ARTICLE{2021ApJ...908..199R,
       author = {{Reichardt}, C.~L. and {Patil}, S. and {Ade}, P.~A.~R. and {Anderson}, A.~J. and {Austermann}, J.~E. and {Avva}, J.~S. and {Baxter}, E. and {Beall}, J.~A. and {Bender}, A.~N. and {Benson}, B.~A. and {Bianchini}, F. and {Bleem}, L.~E. and {Carlstrom}, J.~E. and {Chang}, C.~L. and {Chaubal}, P. and {Chiang}, H.~C. and {Chou}, T.~L. and {Citron}, R. and {Moran}, C. Corbett and {Crawford}, T.~M. and {Crites}, A.~T. and {de Haan}, T. and {Dobbs}, M.~A. and {Everett}, W. and {Gallicchio}, J. and {George}, E.~M. and {Gilbert}, A. and {Gupta}, N. and {Halverson}, N.~W. and {Harrington}, N. and {Henning}, J.~W. and {Hilton}, G.~C. and {Holder}, G.~P. and {Holzapfel}, W.~L. and {Hrubes}, J.~D. and {Huang}, N. and {Hubmayr}, J. and {Irwin}, K.~D. and {Knox}, L. and {Lee}, A.~T. and {Li}, D. and {Lowitz}, A. and {Luong-Van}, D. and {McMahon}, J.~J. and {Mehl}, J. and {Meyer}, S.~S. and {Millea}, M. and {Mocanu}, L.~M. and {Mohr}, J.~J. and {Montgomery}, J. and {Nadolski}, A. and {Natoli}, T. and {Nibarger}, J.~P. and {Noble}, G. and {Novosad}, V. and {Omori}, Y. and {Padin}, S. and {Pryke}, C. and {Ruhl}, J.~E. and {Saliwanchik}, B.~R. and {Sayre}, J.~T. and {Schaffer}, K.~K. and {Shirokoff}, E. and {Sievers}, C. and {Smecher}, G. and {Spieler}, H.~G. and {Staniszewski}, Z. and {Stark}, A.~A. and {Tucker}, C. and {Vanderlinde}, K. and {Veach}, T. and {Vieira}, J.~D. and {Wang}, G. and {Whitehorn}, N. and {Williamson}, R. and {Wu}, W.~L.~K. and {Yefremenko}, V.},
        title = "{An Improved Measurement of the Secondary Cosmic Microwave Background Anisotropies from the SPT-SZ + SPTpol Surveys}",
      journal = {\apj},
         year = 2021,
        month = feb,
       volume = {908},
       number = {2},
          eid = {199},
        pages = {199},
          doi = {10.3847/1538-4357/abd407},
archivePrefix = {arXiv},
       eprint = {2002.06197},
 primaryClass = {astro-ph.CO},
       adsurl = {https://ui.adsabs.harvard.edu/abs/2021ApJ...908..199R}
}

@inproceedings{10.1145/3569951.3597559,
author = {Boerner, Timothy J. and Deems, Stephen and Furlani, Thomas R. and Knuth, Shelley L. and Towns, John},
title = {ACCESS: Advancing Innovation: NSF’s Advanced Cyberinfrastructure Coordination Ecosystem: Services \& Support},
year = {2023},
isbn = {9781450399852},
publisher = {Association for Computing Machinery},
address = {New York, NY, USA},
url = {https://doi.org/10.1145/3569951.3597559},
doi = {10.1145/3569951.3597559},
booktitle = {Practice and Experience in Advanced Research Computing 2023: Computing for the Common Good},
pages = {173–176},
numpages = {4},
location = {Portland, OR, USA},
series = {PEARC '23}
}

@Article{         harris2020array,
 title         = {Array programming with {NumPy}},
 author        = {Charles R. Harris and K. Jarrod Millman and St{\'{e}}fan J.
                 van der Walt and Ralf Gommers and Pauli Virtanen and David
                 Cournapeau and Eric Wieser and Julian Taylor and Sebastian
                 Berg and Nathaniel J. Smith and Robert Kern and Matti Picus
                 and Stephan Hoyer and Marten H. van Kerkwijk and Matthew
                 Brett and Allan Haldane and Jaime Fern{\'{a}}ndez del
                 R{\'{i}}o and Mark Wiebe and Pearu Peterson and Pierre
                 G{\'{e}}rard-Marchant and Kevin Sheppard and Tyler Reddy and
                 Warren Weckesser and Hameer Abbasi and Christoph Gohlke and
                 Travis E. Oliphant},
 year          = {2020},
 month         = sep,
 journal       = {Nature},
 volume        = {585},
 number        = {7825},
 pages         = {357--362},
 doi           = {10.1038/s41586-020-2649-2},
 publisher     = {Springer Science and Business Media {LLC}},
 url           = {https://doi.org/10.1038/s41586-020-2649-2}
}

@Article{Hunter:2007,
  Author    = {Hunter, J. D.},
  Title     = {Matplotlib: A 2D graphics environment},
  Journal   = {Computing in Science \& Engineering},
  Volume    = {9},
  Number    = {3},
  Pages     = {90--95},
  publisher = {IEEE COMPUTER SOC},
  doi       = {10.1109/MCSE.2007.55},
  year      = 2007
}

@article{astropy:2013,
Adsurl = {http://adsabs.harvard.edu/abs/2013A%26A...558A..33A},
Archiveprefix = {arXiv},
Author = {{Astropy Collaboration} and {Robitaille}, T.~P. and {Tollerud}, E.~J. and {Greenfield}, P. and {Droettboom}, M. and {Bray}, E. and {Aldcroft}, T. and {Davis}, M. and {Ginsburg}, A. and {Price-Whelan}, A.~M. and {Kerzendorf}, W.~E. and {Conley}, A. and {Crighton}, N. and {Barbary}, K. and {Muna}, D. and {Ferguson}, H. and {Grollier}, F. and {Parikh}, M.~M. and {Nair}, P.~H. and {Unther}, H.~M. and {Deil}, C. and {Woillez}, J. and {Conseil}, S. and {Kramer}, R. and {Turner}, J.~E.~H. and {Singer}, L. and {Fox}, R. and {Weaver}, B.~A. and {Zabalza}, V. and {Edwards}, Z.~I. and {Azalee Bostroem}, K. and {Burke}, D.~J. and {Casey}, A.~R. and {Crawford}, S.~M. and {Dencheva}, N. and {Ely}, J. and {Jenness}, T. and {Labrie}, K. and {Lim}, P.~L. and {Pierfederici}, F. and {Pontzen}, A. and {Ptak}, A. and {Refsdal}, B. and {Servillat}, M. and {Streicher}, O.},
Doi = {10.1051/0004-6361/201322068},
Eid = {A33},
Eprint = {1307.6212},
Journal = {\aap},
Month = oct,
Pages = {A33},
Primaryclass = {astro-ph.IM},
Title = {{Astropy: A community Python package for astronomy}},
Volume = 558,
Year = 2013}

@ARTICLE{astropy:2018,
       author = {{Astropy Collaboration} and {Price-Whelan}, A.~M. and
         {Sip{\H{o}}cz}, B.~M. and {G{\"u}nther}, H.~M. and {Lim}, P.~L. and
         {Crawford}, S.~M. and {Conseil}, S. and {Shupe}, D.~L. and
         {Craig}, M.~W. and {Dencheva}, N. and {Ginsburg}, A. and {Vand
        erPlas}, J.~T. and {Bradley}, L.~D. and {P{\'e}rez-Su{\'a}rez}, D. and
         {de Val-Borro}, M. and {Aldcroft}, T.~L. and {Cruz}, K.~L. and
         {Robitaille}, T.~P. and {Tollerud}, E.~J. and {Ardelean}, C. and
         {Babej}, T. and {Bach}, Y.~P. and {Bachetti}, M. and {Bakanov}, A.~V. and
         {Bamford}, S.~P. and {Barentsen}, G. and {Barmby}, P. and
         {Baumbach}, A. and {Berry}, K.~L. and {Biscani}, F. and {Boquien}, M. and
         {Bostroem}, K.~A. and {Bouma}, L.~G. and {Brammer}, G.~B. and
         {Bray}, E.~M. and {Breytenbach}, H. and {Buddelmeijer}, H. and
         {Burke}, D.~J. and {Calderone}, G. and {Cano Rodr{\'\i}guez}, J.~L. and
         {Cara}, M. and {Cardoso}, J.~V.~M. and {Cheedella}, S. and {Copin}, Y. and
         {Corrales}, L. and {Crichton}, D. and {D'Avella}, D. and {Deil}, C. and
         {Depagne}, {\'E}. and {Dietrich}, J.~P. and {Donath}, A. and
         {Droettboom}, M. and {Earl}, N. and {Erben}, T. and {Fabbro}, S. and
         {Ferreira}, L.~A. and {Finethy}, T. and {Fox}, R.~T. and
         {Garrison}, L.~H. and {Gibbons}, S.~L.~J. and {Goldstein}, D.~A. and
         {Gommers}, R. and {Greco}, J.~P. and {Greenfield}, P. and
         {Groener}, A.~M. and {Grollier}, F. and {Hagen}, A. and {Hirst}, P. and
         {Homeier}, D. and {Horton}, A.~J. and {Hosseinzadeh}, G. and {Hu}, L. and
         {Hunkeler}, J.~S. and {Ivezi{\'c}}, {\v{Z}}. and {Jain}, A. and
         {Jenness}, T. and {Kanarek}, G. and {Kendrew}, S. and {Kern}, N.~S. and
         {Kerzendorf}, W.~E. and {Khvalko}, A. and {King}, J. and {Kirkby}, D. and
         {Kulkarni}, A.~M. and {Kumar}, A. and {Lee}, A. and {Lenz}, D. and
         {Littlefair}, S.~P. and {Ma}, Z. and {Macleod}, D.~M. and
         {Mastropietro}, M. and {McCully}, C. and {Montagnac}, S. and
         {Morris}, B.~M. and {Mueller}, M. and {Mumford}, S.~J. and {Muna}, D. and
         {Murphy}, N.~A. and {Nelson}, S. and {Nguyen}, G.~H. and
         {Ninan}, J.~P. and {N{\"o}the}, M. and {Ogaz}, S. and {Oh}, S. and
         {Parejko}, J.~K. and {Parley}, N. and {Pascual}, S. and {Patil}, R. and
         {Patil}, A.~A. and {Plunkett}, A.~L. and {Prochaska}, J.~X. and
         {Rastogi}, T. and {Reddy Janga}, V. and {Sabater}, J. and
         {Sakurikar}, P. and {Seifert}, M. and {Sherbert}, L.~E. and
         {Sherwood-Taylor}, H. and {Shih}, A.~Y. and {Sick}, J. and
         {Silbiger}, M.~T. and {Singanamalla}, S. and {Singer}, L.~P. and
         {Sladen}, P.~H. and {Sooley}, K.~A. and {Sornarajah}, S. and
         {Streicher}, O. and {Teuben}, P. and {Thomas}, S.~W. and
         {Tremblay}, G.~R. and {Turner}, J.~E.~H. and {Terr{\'o}n}, V. and
         {van Kerkwijk}, M.~H. and {de la Vega}, A. and {Watkins}, L.~L. and
         {Weaver}, B.~A. and {Whitmore}, J.~B. and {Woillez}, J. and
         {Zabalza}, V. and {Astropy Contributors}},
        title = "{The Astropy Project: Building an Open-science Project and Status of the v2.0 Core Package}",
      journal = {\aj},
         year = 2018,
        month = sep,
       volume = {156},
       number = {3},
          eid = {123},
        pages = {123},
          doi = {10.3847/1538-3881/aabc4f},
archivePrefix = {arXiv},
       eprint = {1801.02634},
 primaryClass = {astro-ph.IM},
       adsurl = {https://ui.adsabs.harvard.edu/abs/2018AJ....156..123A}
}

@ARTICLE{astropy:2022,
       author = {{Astropy Collaboration} and {Price-Whelan}, Adrian M. and {Lim}, Pey Lian and {Earl}, Nicholas and {Starkman}, Nathaniel and {Bradley}, Larry and {Shupe}, David L. and {Patil}, Aarya A. and {Corrales}, Lia and {Brasseur}, C.~E. and {N{"o}the}, Maximilian and {Donath}, Axel and {Tollerud}, Erik and {Morris}, Brett M. and {Ginsburg}, Adam and {Vaher}, Eero and {Weaver}, Benjamin A. and {Tocknell}, James and {Jamieson}, William and {van Kerkwijk}, Marten H. and {Robitaille}, Thomas P. and {Merry}, Bruce and {Bachetti}, Matteo and {G{"u}nther}, H. Moritz and {Aldcroft}, Thomas L. and {Alvarado-Montes}, Jaime A. and {Archibald}, Anne M. and {B{'o}di}, Attila and {Bapat}, Shreyas and {Barentsen}, Geert and {Baz{'a}n}, Juanjo and {Biswas}, Manish and {Boquien}, M{'e}d{'e}ric and {Burke}, D.~J. and {Cara}, Daria and {Cara}, Mihai and {Conroy}, Kyle E. and {Conseil}, Simon and {Craig}, Matthew W. and {Cross}, Robert M. and {Cruz}, Kelle L. and {D'Eugenio}, Francesco and {Dencheva}, Nadia and {Devillepoix}, Hadrien A.~R. and {Dietrich}, J{"o}rg P. and {Eigenbrot}, Arthur Davis and {Erben}, Thomas and {Ferreira}, Leonardo and {Foreman-Mackey}, Daniel and {Fox}, Ryan and {Freij}, Nabil and {Garg}, Suyog and {Geda}, Robel and {Glattly}, Lauren and {Gondhalekar}, Yash and {Gordon}, Karl D. and {Grant}, David and {Greenfield}, Perry and {Groener}, Austen M. and {Guest}, Steve and {Gurovich}, Sebastian and {Handberg}, Rasmus and {Hart}, Akeem and {Hatfield-Dodds}, Zac and {Homeier}, Derek and {Hosseinzadeh}, Griffin and {Jenness}, Tim and {Jones}, Craig K. and {Joseph}, Prajwel and {Kalmbach}, J. Bryce and {Karamehmetoglu}, Emir and {Ka{l}uszy{'n}ski}, Miko{l}aj and {Kelley}, Michael S.~P. and {Kern}, Nicholas and {Kerzendorf}, Wolfgang E. and {Koch}, Eric W. and {Kulumani}, Shankar and {Lee}, Antony and {Ly}, Chun and {Ma}, Zhiyuan and {MacBride}, Conor and {Maljaars}, Jakob M. and {Muna}, Demitri and {Murphy}, N.~A. and {Norman}, Henrik and {O'Steen}, Richard and {Oman}, Kyle A. and {Pacifici}, Camilla and {Pascual}, Sergio and {Pascual-Granado}, J. and {Patil}, Rohit R. and {Perren}, Gabriel I. and {Pickering}, Timothy E. and {Rastogi}, Tanuj and {Roulston}, Benjamin R. and {Ryan}, Daniel F. and {Rykoff}, Eli S. and {Sabater}, Jose and {Sakurikar}, Parikshit and {Salgado}, Jes{'u}s and {Sanghi}, Aniket and {Saunders}, Nicholas and {Savchenko}, Volodymyr and {Schwardt}, Ludwig and {Seifert-Eckert}, Michael and {Shih}, Albert Y. and {Jain}, Anany Shrey and {Shukla}, Gyanendra and {Sick}, Jonathan and {Simpson}, Chris and {Singanamalla}, Sudheesh and {Singer}, Leo P. and {Singhal}, Jaladh and {Sinha}, Manodeep and {Sip{H{o}}cz}, Brigitta M. and {Spitler}, Lee R. and {Stansby}, David and {Streicher}, Ole and {{{S}}umak}, Jani and {Swinbank}, John D. and {Taranu}, Dan S. and {Tewary}, Nikita and {Tremblay}, Grant R. and {Val-Borro}, Miguel de and {Van Kooten}, Samuel J. and {Vasovi{'c}}, Zlatan and {Verma}, Shresth and {de Miranda Cardoso}, Jos{'e} Vin{'i}cius and {Williams}, Peter K.~G. and {Wilson}, Tom J. and {Winkel}, Benjamin and {Wood-Vasey}, W.~M. and {Xue}, Rui and {Yoachim}, Peter and {Zhang}, Chen and {Zonca}, Andrea and {Astropy Project Contributors}},
        title = "{The Astropy Project: Sustaining and Growing a Community-oriented Open-source Project and the Latest Major Release (v5.0) of the Core Package}",
      journal = {\apj},
         year = 2022,
        month = aug,
       volume = {935},
       number = {2},
          eid = {167},
        pages = {167},
          doi = {10.3847/1538-4357/ac7c74},
archivePrefix = {arXiv},
       eprint = {2206.14220},
 primaryClass = {astro-ph.IM},
       adsurl = {https://ui.adsabs.harvard.edu/abs/2022ApJ...935..167A}
}

@ARTICLE{2019arXiv191201703P,
       author = {{Paszke}, Adam and {Gross}, Sam and {Massa}, Francisco and {Lerer}, Adam and {Bradbury}, James and {Chanan}, Gregory and {Killeen}, Trevor and {Lin}, Zeming and {Gimelshein}, Natalia and {Antiga}, Luca and {Desmaison}, Alban and {K{\"o}pf}, Andreas and {Yang}, Edward and {DeVito}, Zach and {Raison}, Martin and {Tejani}, Alykhan and {Chilamkurthy}, Sasank and {Steiner}, Benoit and {Fang}, Lu and {Bai}, Junjie and {Chintala}, Soumith},
        title = "{PyTorch: An Imperative Style, High-Performance Deep Learning Library}",
      journal = {arXiv e-prints},
         year = 2019,
        month = dec,
          eid = {arXiv:1912.01703},
        pages = {arXiv:1912.01703},
          doi = {10.48550/arXiv.1912.01703},
archivePrefix = {arXiv},
       eprint = {1912.01703},
 primaryClass = {cs.LG},
       adsurl = {https://ui.adsabs.harvard.edu/abs/2019arXiv191201703P}
}

@ARTICLE{sundararajan2017axiomatic,
  title={Axiomatic attribution for deep networks},
  author={Sundararajan, Mukund and Taly, Ankur and Yan, Qiqi},
  journal={International conference on machine learning},
  pages={3319--3328},
  year={2017},
  organization={PMLR}
}

@article{smilkov2017smoothgrad,
  title={Smoothgrad: removing noise by adding noise},
  author={Smilkov, Daniel and Thorat, Nikhil and Kim, Been and Vi{\'e}gas, Fernanda and Wattenberg, Martin},
  journal={arXiv preprint arXiv:1706.03825},
  year={2017}
}
\bibliographystyle{aasjournal}



\end{document}